\documentclass[11pt]{article}
\usepackage{graphicx}
\usepackage{amsfonts}
\usepackage{amsmath}
\usepackage{amssymb}
\usepackage{url}
\usepackage{graphicx}
\usepackage{tikz}
\usepackage{fancyhdr}
\usetikzlibrary{calc}
\usetikzlibrary{positioning}
\usetikzlibrary{arrows}
\usepackage{float}
\usepackage{subcaption}
\usepackage{hhline}
\usepackage[title]{appendix}

\usepackage{indentfirst}
\usepackage{enumerate}
\usepackage{amsthm}
\usepackage{color}
\usepackage{comment}
\usepackage{natbib}
\usepackage{dsfont}
\usepackage[colorlinks=true,citecolor=blue]{hyperref}

\usepackage[misc]{ifsym}

\usepackage{booktabs}
\usepackage{amssymb}
\usepackage{graphicx}   
\usepackage{geometry}
\usepackage{subcaption}

\tikzset{
	block/.style={rectangle, draw, rounded corners, text centered, text width = 7em, minimum height = 2em},
	line/.style={draw, -latex'}
}

\def\d{\,\mathrm{d}}
\def\laweq{\buildrel \d \over =}

\newcommand{\VaR}{\mathrm{VaR}}

\newcommand{\KL}{\mathrm{KL}}

\newcommand{\ES}{\mathrm{ES}}
\newcommand{\E}{\mathbb{E}}
\newcommand{\R}{\mathbb{R}}

\newcommand{\id}{\mathds{1}}

\newcommand{\X}{\mathcal X}
\renewcommand{\P}{\mathcal P}

\newcommand{\Q}{\mathcal Q}

\renewcommand{\ge}{\geqslant}
\renewcommand{\le}{\leqslant}
\renewcommand{\geq}{\geqslant}
\renewcommand{\leq}{\leqslant}
\renewcommand{\epsilon}{\varepsilon}
\theoremstyle{plain}
\newtheorem{theorem}{Theorem}
\newtheorem{corollary}{Corollary}
\newtheorem{lemma}{Lemma}
\newtheorem{proposition}{Proposition}

\theoremstyle{definition}
\newtheorem{definition}{Definition} 
\newtheorem{assumption}{Assumption} 
 
\theoremstyle{remark}
\newtheorem{remark}{Remark} 
\theoremstyle{definition}

\renewcommand{\cite}{\citet} 

\usepackage{setspace}

\begin{document}
\title{
Diversified-Risk-Perception Risk Measures and Portfolio Selection
}

\author{Yang Liu\thanks{\scriptsize School of Science and Engineering, The Chinese University of Hong Kong (Shenzhen), Shenzhen, China. Email: \texttt{yangliu16@cuhk.edu.cn}}
\and
Yuhao Liu\thanks{\scriptsize Corresponding Author. School of Science and Engineering, The Chinese University of Hong Kong (Shenzhen), Shenzhen, China. Email: \texttt{liuyuhao@cuhk.edu.cn}}
\and
Jiani Shen\thanks{\scriptsize  Fullgoal Fund Management Co., Ltd., Shanghai, China. Email: \texttt{jianish@126.com}}
}

\date{}
\maketitle

\begin{abstract}

    Traditional risk measures generally rely on a single probability model, potentially overlooking information from alternative models and leading to biased risk assessments. This limitation cannot be addressed by asset diversification alone in risk management. In this paper, we propose a Diversified-Risk-Perception (DRP) risk measure framework that integrates asset diversification and risk model diversification to provide a more comprehensive assessment of market risk while reducing reliance on any single model. First, we introduce a flexible exponential aggregation function to evaluate the risk of a portfolio, which accounts for the contributions of all loss components while placing greater emphasis on large losses. Second, we incorporate multiple risk models in the risk measure and optimize their weights based on  their perceptions on the risk assessments. As a specific example, we introduce DRP expected shortfall (DRP-ES),  establish its theoretical properties, and derive bounds quantifying its sensitivity to model parameters. Empirical experiments in the U.S., Chinese A-share, and South Korean stock markets show that trading strategies based on the DRP-ES achieve higher Sharpe ratios and lower maximum drawdowns than market benchmarks across different periods. Compared with asset diversification strategies based on individual probability models, the DRP strategy exhibits more stable and robust performance across different markets and periods. These findings suggest that the DRP framework can adapt to changing market conditions and improve the stability and effectiveness of portfolio risk management.

	~
	
\noindent \begin{bfseries}Keywords\end{bfseries}:  
Risk management; 
Portfolio optimization; 
Diversification;  Exponential aggregation function; Diversified-Risk-Perception expected shortfall (DRP-ES).	

\end{abstract}


\newpage

\section{Introduction}

As global economic uncertainty increases, financial institutions are compelled to improve their risk measurement methods to make better asset allocation decisions.
Empirical evidence supports this necessity. In 2020, the COVID-19 pandemic caused a liquidity shortage and the U.S. stock market triggered circuit breakers four times within ten days. In 2022, the Ukraine crisis  led to sharp increases in energy and commodity prices, adding to inflationary pressure and increasing uncertainty in global financial markets. 
In 2024, the Federal Reserve maintained high interest rates, raising borrowing costs and putting pressure on emerging-market currencies and capital flows, thereby increasing uncertainty in global financial markets.
In 2025, tariff shocks pushed the VIX above 60, indicating extreme market volatility. In 2026, the geopolitical crisis in the Middle East disrupted energy and non-ferrous metal markets, leading to increased volatility. Moreover, in the South Korean stock market, the benchmark KOSPI index plunged as corporate earnings fell short of expectations and brokers forcibly liquidated many loss-making positions,
putting the market on track for its steepest monthly decline on record.  
These events show that financial risks arise from diverse sources and involve considerable uncertainty, posing significant challenges to traditional risk management paradigms that rely on a single risk model.


Investors commonly rely on asset diversification to manage financial risk, while often overlooking risk model diversification. Although asset diversification can reduce asset-specific
risk by spreading investments across different assets, it cannot resolve uncertainty in risk estimation. Traditional risk measures generally rely on a single probability model to assess financial losses. However, relying on a single selected model can ignore model uncertainty and lead to overconfident inferences (see \cite{hoeting1999bayesian}). Different risk models may capture different features of the loss distribution and consequently produce different risk estimates even for the same portfolio. \cite{danielsson2016model} documented substantial disagreement among risk forecasts from different risk models, particularly during periods of market distress. Hence, integrating information from multiple risk models offers a way to reduce reliance on a single specification and improve the effectiveness of risk estimation (see \cite{opschoor2017combining} and \cite{taylor2020forecast}). These findings motivate a framework that not only accounts for asset diversification but also considers risk model diversification.


A natural starting point for diversifying risk models is to consider how risk measures can incorporate information from multiple probability models. 
Several studies have considered risk assessment under multiple probability models (see \cite{wang2021scenario} and \cite{FLW21}).
\cite{FLW21} developed the concept of generalized risk measures (GRM) and its axiomatic characterization. Unlike the classical definition of risk measures in \cite{ADEH99} and \cite{FS16}, 
in the GRM framework, the input of a risk measure  is a combination of the loss random variables and a set of possible probability measures that represents the best knowledge of the underlying random nature. Although GRM provides a general theoretical framework, its practical implementation in risk management remains challenging. In particular, the framework does not directly specify how to combine information from different probability models or how to assign appropriate weights to them.  \cite{FLW21} studied worst-case generalized risk measures, which evaluate risk by taking the highest risk estimate among the candidate probability models. \cite{zhu2009worst} considered the worst-case Conditional Value-at-Risk for robust portfolio optimization, while \cite{zymler2013worst} studied the worst-case VaR approximations for derivative portfolios. These approaches provide a conservative risk assessment and help avoid underestimating potential losses under model uncertainty.  However, since the risk estimate is determined by the most pessimistic model, it may not fully reflect the information provided by other probability models. The thick modelling approach in \cite{pesaran2009model} assigns equal weights to a selected subset of models, potentially overlooking useful information from excluded models and underestimating tail risk by assigning insufficient weights to models that capture extreme losses. These limitations motivate the development of a practical method that systematically integrates multiple probability models to provide more reliable risk assessments.


While diversifying risk models allows risk to be assessed from different probabilistic perspectives, asset diversification aims to manage the risk by spreading investments across different assets. To assess the risk of a diversified portfolio, we need to consider the risk aggregation (see \cite{chen2022ordering} and \cite{blanchet2025convolution}). Aggregation functions provide a way to combine a vector of losses into a scalar loss to which a univariate risk measure can be applied (see \cite{burgert2006consistent}). As for the form of the aggregation function,
\cite{chen2013axiomatic}, \cite{muller2023description} and \cite{de2026upper} discussed the linear weighted aggregation function, while \cite{lux2019model} considered the maximum aggregation function. However, the linear weighted aggregation function may give insufficient attention to individual assets with large losses, while the maximum aggregation function only considers the largest loss and ignores the contributions of other assets. Therefore, neither function alone provides a flexible balance between accounting for all losses and emphasizing large losses. This motivates the use of a more flexible aggregation function that gives greater attention to large losses without ignoring the contributions of other assets.

In this paper, we propose a Diversified-Risk-Perception (DRP) risk measure framework that combines  risk model diversification and asset diversification by considering random loss vectors together with multiple candidate probability models. First, we introduce a flexible aggregation function that includes the linear weighted aggregation function and the maximum aggregation function as limiting cases. This allows us to account for all loss components while giving greater attention to large losses. The form of the aggregation function is determined by the historical losses of the stocks in each portfolio, which allows the aggregation function to reflect the risk characteristics of different portfolios and adjust its emphasis on large losses. Second, we construct a mixture probability model and optimize the model weights based on the level of disagreement among the risk forecasts produced by the candidate probability models. When the model disagreement is small, the weights of different models tend to be similar, while greater disagreement allows more weight to be assigned to pessimistic models to avoid underestimating risk, without necessarily excluding the information provided by models that predict smaller losses. As a specific example of the DRP risk measure framework, we analyze the properties of the DRP expected shortfall (DRP-ES) and
derive an upper bound on the change in the DRP-ES, as model disagreement and the confidence level vary. Numerical experiments in the U.S., Chinese A-share, and South Korean stock markets show that the trading strategy based on the DRP-ES achieves higher Sharpe ratios and lower maximum drawdowns than the market benchmarks across different periods, while exhibiting more stable performance than strategies based on individual probability models. These results demonstrate the benefits of combining flexible loss aggregation with information from multiple probability models for portfolio risk management.

Our contributions can be summarized as follows. First, we propose a DRP risk measure framework that combines asset diversification with the risk model diversification. We introduce a flexible non-linear aggregation function for evaluating the risk of a portfolio and develop a method for combining multiple candidate probability models that assess risk from different perspectives. Second, we establish the theoretical properties of the DRP-ES, including its convexity and cash invariance. We also derive an upper bound on the change in the DRP-ES as model parameters vary and characterize the limiting behavior of the optimal model weights. Third, we verify the effectiveness and robustness of the DRP risk measure framework in three financial markets and show that combining asset diversification with risk model diversification provides more stable performance than asset diversification alone in risk management.

The remainder of this paper is organized as follows. Section \ref{sec:general_drp} introduces the general Diversified-Risk-Perception risk measure framework, including the aggregation of vector losses and the integration of multiple probability models. Section \ref{sec:DRP-ES} defines the DRP-ES, analyzes the limiting behavior of optimal model weights, and presents its numerical computation method. Section \ref{sec:DRP-ES-analysis} establishes the theoretical properties of the DRP-ES and analyzes its sensitivity to model parameters. Section \ref{sec:Numerical Experiments} conducts numerical experiments in the U.S., Chinese A-share, and South Korean stock markets to evaluate the performance of the proposed trading strategies based on the DRP-ES and the single-scenario expected shortfall. Section \ref{sec:conclusion} concludes the paper. Proofs are included in Appendix \ref{app:proof}.

\section{Diversified-Risk-Perception Risk Measures}
\label{sec:general_drp}
In this section, we introduce a general class of
diversified-risk-perception (DRP) risk measures.
Let $(\Omega,\mathcal F)$ be a measurable space, let $\mathcal P$ be a collection
of probability measures on $(\Omega,\mathcal F)$, and let $\mathcal X$ be a
linear space of real-valued measurable random variables containing the
constants. We denote by
\[
\mathcal X^d=\left\{\mathbf X=(X_1,\ldots,X_d):
X_i\in\mathcal X,\ i=1,\ldots,d\right\}
\]
the space of $d$-dimensional random loss vectors. We provide the definition of the generalized risk measure (GRM) in \cite{FLW21}:
\begin{definition}\label{def:GRM}
	A  \emph{generalized risk measure} is a mapping $\Psi:\X\times 2^\P \to [-\infty,\infty]$. The generalized risk measure $\Psi$ is called \emph{standard} if ${\Psi(s\mid \Q) = s}$ for all ${s\in \R}$ and ${\Q\subseteq \mathcal P}$.
	A \emph{single-scenario risk measure} is a mapping $\Psi:\X \times \P\to [-\infty,\infty]$. 
\end{definition}
Following the generalized-risk-measure framework of
\cite{FLW21}, let $\Gamma:\mathcal X\times\mathcal P\longrightarrow\mathbb R$
be a single-scenario risk measure, where $\Gamma(Y\mid P)$
denotes the risk associated with the scalar loss $Y\in\mathcal X$ under
the probability measure $P\in\mathcal P$. For example, for a level $\alpha\in(0,1)$, the value-at-risk (VaR) under $P$ is defined as $$\mathrm{VaR}_\alpha(X\mid P) = \inf\{x \in \mathbb{R} : P[X \leq x] \geq \alpha\}, \quad X \in \mathcal{X},$$
and the expected shortfall (ES) under $P$ is defined as
\[
\mathrm{ES}_\alpha(X|P) = \frac{1}{1-\alpha} \int_\alpha^1 \mathrm{VaR}_\beta(X\mid P) \, \mathrm{d}\beta, \qquad X \in \mathcal{X}.
\]

Definition \ref{def:GRM} provides a general framework for measuring scalar losses under a single probability measure or a set of probability models. In practice, a portfolio usually consists of multiple assets, each of which may contribute differently to the overall risk. Moreover, although using multiple probability models allows us to assess risk from different perspectives, determining how to assign appropriate weights to these probability models remains an important challenge, as inappropriate model weights may lead to unreliable risk assessments. In general, when the probability models exhibit substantial disagreement in their assessments of potential losses, we prefer the models that provide more conservative risk assessments to avoid underestimating the tail risk and potential large losses. Conversely, when the probability models have similar assessments of potential losses, there is no clear preference for any of them.
To address these aspects, we introduce two components. First, we introduce a flexible non-linear aggregation function that transforms a vector of losses into a scalar loss in Section \ref{sec:aggregation-risk-vector}. Second, we quantify the disagreement among the candidate probability models in their risk assessments to determine the weights assigned to each model in Section \ref{sec:model-mixtures}. By combining these two components, we obtain the proposed DRP risk measure. Throughout the paper, we denote by $||\cdot||_1$ the $\ell^1$-norm of a vector.

\subsection{Aggregation of Losses}\label{sec:aggregation-risk-vector}

To accommodate multidimensional risks, let $\Lambda_\theta:\mathbb R^d\longrightarrow\mathbb R$
be a deterministic Borel-measurable aggregation function, where
$\theta$ denotes a parameter set. For
$\mathbf X\in\mathcal X^d$, we define
\[
    \Lambda_\theta(\mathbf X)(\omega)
    :=
    \Lambda_\theta
    \bigl(
        X_1(\omega),\ldots,X_d(\omega)
    \bigr).
\]
We consider the following properties of the aggregation function.
\begin{enumerate}
    \item[(L0)] \emph{Constant preservation:}
    \[
        \Lambda_\theta(c\mathbf 1_d)=c,
        \qquad c\in\mathbb R,
    \]
    where $\mathbf 1_d=(1,\ldots,1)^\top$.

    \item[(L1)] \emph{Componentwise monotonicity:}
    \[
        \mathbf X\leq\mathbf Y
        \quad\Longrightarrow\quad
        \Lambda_\theta(\mathbf X)
        \leq
        \Lambda_\theta(\mathbf Y).
    \]

    \item[(L2)] \emph{Cash invariance:}
    \[
        \Lambda_\theta(\mathbf X+c\mathbf 1_d)
        =
        \Lambda_{\theta}(\mathbf X)+c,
        \qquad c\in\mathbb R.
    \]

    \item[(L3)] \emph{Convexity:}
    \[
        \Lambda_\theta
        \bigl(
            \lambda\mathbf X+(1-\lambda)\mathbf Y
        \bigr)
        \leq
        \lambda\Lambda_{\theta}(\mathbf X)
        +(1-\lambda)\Lambda_\theta(\mathbf Y),
        \qquad \lambda\in[0,1].
    \]

    \item[(L4)] \emph{Positive homogeneity:}
    \[
        \Lambda_\theta(\lambda\mathbf X)
        =
        \lambda\Lambda_{\theta}(\mathbf X),
        \qquad \lambda\geq0.
    \]

    \item[(L5)] \emph{Subadditivity:}
    \[
        \Lambda_\theta(\mathbf X+\mathbf Y)
        \leq
        \Lambda_\theta(\mathbf X)
        +
        \Lambda_\theta(\mathbf Y).
    \]
\end{enumerate}

Condition (L0) guarantees consistency with deterministic losses, whereas
(L1) and (L2) provide the natural compatibility conditions between the
multivariate aggregation and the monetary-risk-measure structure.
Conditions (L3)--(L5) determine whether convexity or coherence can be
inherited by the resulting vector-input risk measure. 

There are some typical aggregation functions: 
\begin{itemize}
    \item The linear weighted aggregation function (\cite{de2026upper}): 
    \[
    \Lambda_\theta(\mathbf X)=\boldsymbol{\pi}^\top \mathbf X=\sum_{i=1}^d\pi_iX_i, \quad \pi_i\in [0,\infty).
    \]
    \item The maximum aggregation function (\cite{lux2019model}):
    \[
    \Lambda_\theta(\mathbf X)=\max_{1\leq i\leq d}X_i.
    \]
\end{itemize}
If \(\sum_{i=1}^d\pi_i=1\), then both the linear weighted aggregation function and the maximum aggregation function satisfy all the properties (L0)-(L5).
The linear weighted aggregation function and the maximum aggregation function have distinct features. To incorporate both within a single framework, we adopt the exponential aggregation function:
\begin{align}\label{eq:normalized-log-sum-exp-aggregation}
    F_{\theta,\boldsymbol{\pi}}(\mathbf X)
    =
    \theta
    \log
    \left(
        \sum_{i=1}^d \pi_i
        e^{X_i/\theta}
    \right),
\end{align}
where $\theta>0$, $\boldsymbol\pi=(\pi_1,\cdots,\pi_d)^\top$ and $\sum_{i=1}^d\pi_i=1$. 
We do not allow short-selling, hence we let $\pi_i\in (0,\infty)$.
Aggregation function \eqref{eq:normalized-log-sum-exp-aggregation} satisfies (L0)-(L3), but does not satisfy (L4) and (L5) in general. In Proposition \ref{prop:Limiting-relations-of-log-sum-exp-aggregations}, we  show that the exponential aggregation function \eqref{eq:normalized-log-sum-exp-aggregation} approaches the maximum aggregation function as \(\theta\) tends to zero and the linear weighted aggregation function as \(\theta\) tends to infinity.
\begin{proposition}\label{prop:Limiting-relations-of-log-sum-exp-aggregations}
For $\mathbf X=(X_1,\ldots,X_d)$, we have
\begin{align*}
    \lim_{\theta\downarrow0}
    F_{\theta,\boldsymbol{\pi}}(\mathbf X)=
    \max_{1\leq i\leq d}X_i, \qquad
    \lim_{\theta\to\infty}
    F_{\theta,\boldsymbol{\pi}}(\mathbf X)
    =\boldsymbol{\pi}^\top\mathbf X.
\end{align*}
\end{proposition}
\begin{proof}
    See Appendix \ref{app:proof}.
\end{proof}


\subsection{Probability Models}\label{sec:model-mixtures}
In practice, different financial institutions may use different probability models to assess the same potential losses, as they rely on different data sources, modeling assumptions, and estimation methods. These models capture different features of the loss distribution and therefore produce different risk assessments. By integrating the risk assessments of different probability models, we can obtain a more comprehensive evaluation of the risk than relying on a single source.

At each time $t$, suppose that there are $M$ candidate probability models: $P_{1,t},\ldots,P_{M,t}\in\mathcal P.$
We impose the following condition for the probability models:
\begin{assumption}
\label{ass:bounded-aggregated-loss}
Let \(\alpha\in(0,1)\). For any random loss vector $\mathbf X\in \mathcal X^d$ and any aggregation function $\Lambda_\theta(\mathbf X)$, there exist finite constants
\(\underline y\le \overline y\) such that
\[
P_{m,t}\left(\underline y\le
\Lambda_\theta(\mathbf X)\le \overline y\right)=1,\qquad
m=1,\ldots,M.
\]
\end{assumption}

Let
\[
    \Delta_M
    :=
    \left\{
        \mathbf q=(q_1,\ldots,q_M)^\top\in\mathbb R_+^M:
        \sum_{m=1}^M q_m=1
    \right\}.
\]
For $\mathbf q\in\Delta_M$, we define the mixture probability model
\begin{equation}
    P_t^{\mathbf q}=
    \sum_{m=1}^M q_m P_{m,t}.
    \label{eq:mixture_probability}
\end{equation}
We assume that $P_t^{\mathbf q}\in \mathcal P$. Let $\boldsymbol\omega_0=(\omega_{0,1},\ldots,\omega_{0,M})^\top\in\Delta_M$
be an initial weight vector of different models. Throughout the paper, we set $\boldsymbol\omega_0=\left(\frac1M,\ldots,\frac1M\right)^\top$. 
For $\mathbf q\in\Delta_M$, we define the Kullback--Leibler divergence
from $\mathbf q$ to $\boldsymbol\omega_0$ by
\begin{equation}
    \operatorname{KL}
    (\mathbf q\mid\boldsymbol\omega_0)
    :=
    \sum_{m=1}^M
    q_m
    \log
    \frac{q_m}{\omega_{0,m}},
    \label{eq:KL_general}
\end{equation}
with the convention $0\log0=0$.
\begin{lemma}\label{lem:upper-bound-of-KL-radius}
    For any $\mathbf q \in \Delta_M,$ we have
    \[
    \operatorname {KL}(\mathbf q \mid \boldsymbol\omega_0)\le \log M.
    \]  
\end{lemma}
\begin{proof}[Proof of Lemma \ref{lem:upper-bound-of-KL-radius}]
For any \(\mathbf q \in \Delta_M\), we have
\[
\begin{aligned}
\operatorname {KL}(\mathbf q \mid \boldsymbol\omega_0)
=
\sum_{m=1}^M
q_m\log\frac{q_m}{1/M}=
\log M+\sum_{m=1}^Mq_m\log q_m\le
\log M.
\end{aligned}
\]
\end{proof}

Different probability models may produce different assessments of potential losses. Some models may predict larger losses, while others may predict smaller losses, leading to disagreement among the models. When the disagreement is substantial, we prefer to assign greater weights to more pessimistic models to avoid underestimating risk. Conversely, when the disagreement is small, the models provide similar risk assessments, and we prefer to assign approximately equal weights to all models.
Let $D_t\geq0$ quantify the level of disagreement among the candidate
models at time $t$. 
We define the ambiguity radius by
\begin{equation}
    r(D_t)
    :=
    \delta_0+\delta_1D_t,
    \qquad
    \delta_0\geq0,\quad \delta_1>0,
    \label{eq:ambiguity_radius}
\end{equation}
and the corresponding ambiguity set of model weights by
\begin{equation}
    \mathcal U_t(D_t)
    :=
    \left\{
        \mathbf q\in\Delta_M:
        \operatorname{KL}
        (\mathbf q\mid\boldsymbol\omega_0)
        \leq
        \delta_0+\delta_1D_t
    \right\}.
    \label{eq:general_ambiguity_set}
\end{equation}
The parameter $\delta_0$ improves robustness by preventing the risk measure from relying too heavily on the model weights \(\boldsymbol\omega_0\) even when the candidate models exhibit no disagreement,
whereas the parameter \(\delta_1\) controls how strongly model disagreement affects the risk measure.

\section{Diversified-Risk-Perception Expected Shortfall}\label{sec:DRP-ES}
The general DRP risk measure can be defined as
\begin{align}\label{eq:general-DRP-risk-measure}
\sup_{\mathbf q\in\mathcal{U}_t(D_t)}\Gamma\left(\Lambda_{\theta}(\mathbf X)\mid P_t^{\mathbf q}\right).
\end{align}
Hence, the DRP expected shortfall (DRP-ES) is defined as follows:
\begin{align}\label{eq:diversified-risk-perception-risk-measure}
    \Psi_{t,\alpha}(D, \mathbf X) := \sup_{\mathbf q\in\mathcal{U}_t(D)}\operatorname{ES}_{\alpha}\left(\Lambda_{\theta}(\mathbf X)\mid P_t^{\mathbf q}\right),
\end{align}
where $\Lambda_{\theta}$ is the aggregation function. 
\begin{theorem}
\label{thm:limiting_optimal_weights}
For any random loss vector $\mathbf X\in \mathcal X^d$ and any aggregation function $\Lambda_\theta$, let $Y=\Lambda_{\theta}(\mathbf X)$. 
For each \(D\geq0\), assume that
\[
    \mathbf q_D^*
    =
    \arg\sup_{\mathbf q\in\mathcal U_t(D)}
    \operatorname{ES}_{\alpha}
    \left(
        Y\mid P_t^{\mathbf q}
    \right).
\]
Suppose that there exists a candidate model \(m^*\) and a threshold
\(z_\alpha^*\in \mathbb R\) satisfying
\[
    P_{m^*,t}(Y<z_\alpha^*)
    \leq \alpha
    \leq
    P_{m^*,t}(Y\leq z_\alpha^*),
\]
such that
\[
    \mathbb E_{P_{m^*,t}}
    \left[(Y-z_\alpha^*)_+\right]
    >
    \mathbb E_{P_{m,t}}
    \left[(Y-z_\alpha^*)_+\right]
\]
for every
\(m\in\{1,\ldots,M\}\setminus\{m^*\}\), where $(Y-z_\alpha^*)_+=\max(Y-z_\alpha^*,0).$
Then the following statements hold.

\begin{enumerate}
    \item[(i)]
As \(D\downarrow0\) and \(\delta_0\downarrow0\), we have
\[
    \mathbf q_{D}^*
    \longrightarrow
    \boldsymbol{\omega}_0
    =
    \left(
        \frac{1}{M},\ldots,\frac{1}{M}
    \right)^\top.
\]

    \item[(ii)]
    As \(D\to\infty\), we have
    \[
        \mathbf q_D^*
        \longrightarrow
        \mathbf e_{m^*},
    \]
    where \(\mathbf e_{m^*}\) assigns unit weight to the model \(m^*\) and zero weight to all other models.
\end{enumerate}
\end{theorem}
\begin{proof}
    See Appendix \ref{app:proof}.
\end{proof}

Theorem \ref{thm:limiting_optimal_weights} shows that, when \(\delta_0\) is small and the candidate models provide very similar assessments of the risk, the mixture probability model tends to assign similar weights to the different candidate models, thereby making approximately equal use of the information provided by each model. In contrast, when the candidate models exhibit substantial disagreement in their risk assessments, the mixture probability model tends to assign the largest weight to the most pessimistic model in terms of the tail risk under certain conditions, meaning that the DRP-ES converges to the worst-case ES (\cite{FLW21}) if the disagreement between the models is large enough.
\begin{remark}
    If \(M=1\), the DRP-ES reduces to the single-scenario ES, which is computed as
\[
\operatorname{ES}_{\alpha}\left(\Lambda_{\theta}(\mathbf X)\mid P_{1,t}\right).
\]
\end{remark}

\subsection{Numerical Computation of DRP-ES}
At each time step $t$, we generate $N$ scenarios of the
$d$-dimensional loss vector $\mathbf X$ under each candidate
probability model $P_{m,t}$. We denote the $n$-th scenario under
model $m$ by
\[
\mathbf X_{m,t}^{(n)}
=
\left(
X_{m,t,1}^{(n)},\ldots,X_{m,t,d}^{(n)}
\right)^\top
\in\mathbb R^d,
\qquad
m=1,\ldots,M,\quad n=1,\ldots,N,
\]
where $X_{m,t,i}^{(n)}$ is the loss of asset $i$ in that scenario. We apply the aggregation function \eqref{eq:normalized-log-sum-exp-aggregation} to each
scenario separately and obtain the scalar aggregated losses
\[
Y_{m,t}^{(n)}
=
\Lambda_\theta\!\left(\mathbf X_{m,t}^{(n)}\right)
=
\theta\log\!\left(
\sum_{i=1}^{d}
\pi_i\exp\!\left\{
\frac{X_{m,t,i}^{(n)}}{\theta}
\right\}
\right),
\qquad
m=1,\ldots,M,\quad n=1,\ldots,N.
\]
Here $\theta>0$ and the aggregation weights $\pi_i$ are as
defined in \eqref{eq:normalized-log-sum-exp-aggregation}. This construction yields $N$ scalar loss
scenarios for each model.

To measure the disagreement among the $M$ candidate models, we use the
pairwise Wasserstein distance between their simulated loss distributions.
For each model $m$, let
\[
    Y_{m,t}^{(1:N)}
    \leq
    Y_{m,t}^{(2:N)}
    \leq \cdots \leq
    Y_{m,t}^{(N:N)}
\]
denote the order statistics. Since all candidate models generate the same
number of scenarios, the empirical one-dimensional Wasserstein distance
between models $m$ and $\ell$ is
\[
    W_{m\ell,t}
    =
    \frac{1}{N}
    \sum_{n=1}^{N}
    \left|
        Y_{m,t}^{(n:N)}
        -
        Y_{\ell,t}^{(n:N)}
    \right|.
\]
Then we average the Wasserstein distances over all pairs of models:
\[
    \overline W_t=
    \frac{2}{M(M-1)}
    \sum_{1\leq m<\ell\leq M}
    W_{m\ell,t}.
\]
We normalize the average of the Wasserstein distances by the standard deviation of all simulated losses. Specifically,
let
\[
    \overline Y_t=
    \frac{1}{MN}
    \sum_{m=1}^{M}\sum_{n=1}^{N}Y_{m,t}^{(n)}
\]
and
\[
    s_t=
    \left[
        \frac{1}{MN}
        \sum_{m=1}^{M}\sum_{n=1}^{N}
        \left(Y_{m,t}^{(n)}-\overline Y_t\right)^2
    \right]^{1/2}.
\]
The model disagreement used in the ambiguity set is then defined as
\[
    D_t=
    \frac{\overline W_t}{s_t+\varepsilon},
\]
where $\varepsilon>0$ is a small  constant used to avoid division
by zero. A larger $D_t$ indicates greater disagreement among the candidate
models and leads to a larger
ambiguity set of model weights through $r(D_t)=\delta_0+\delta_1D_t$.

By the  dual representation of the expected shortfall, we have 
\begin{align}
\ES_\alpha(Y) = \max_{\mathbf u \in \mathbb{R}^N} \left\{ \sum_{i=1}^N u_i y_i : 0 \le u_i \le \frac{p_i}{1-\alpha}, \quad \sum_{i=1}^N u_i = 1 \right\},
\end{align}
where $\mathbf u=(u_1,u_2,\cdots,u_N)$, $y_i$ is the loss in scenario $i$ and $p_i$ is the corresponding probability. For the model $m$, we have $p_i=\frac{q_m}{N}$, where $q_m$ is the weight for the model $m$. So the empirical approximation to the DRP-ES \eqref{eq:diversified-risk-perception-risk-measure} can be obtained by solving the convex optimization problem below with standard numerical methods:
\begin{align}
\max_{\mathbf q, \mathbf u} \quad & \sum_{m=1}^{M} \sum_{n=1}^{N} u_{m,n} Y_{m,t}^{(n)}\nonumber \\
\text{s.t.} \quad & \mathbf q \in \Delta_M, \nonumber \\
& \KL(\mathbf q \mid \boldsymbol \omega_0) \le \delta_0 + \delta_1 D_t, \\
& \sum_{m=1}^{M} \sum_{n=1}^{N} u_{m,n} = 1, \nonumber\\
& 0 \le u_{m,n} \le \frac{q_m}{N(1 - \alpha)}, \quad \forall m, n. \nonumber
\end{align}

\section{Theoretical Analysis of DRP-ES}
\label{sec:DRP-ES-analysis}
\subsection{Sensitivity Analysis}
In this section, we investigate how the DRP-ES
responds to changes in the level of disagreement among candidate probability models and the confidence level $\alpha$. For two probability measures \(\mu\) and \(\nu\) on the same measurable space, define 
\[d_{\mathrm{TV}}(\mu,\nu)=\sup_{A}|\mu(A)-\nu(A)|.\]
We introduce the following lemmas for the proof of the sensitivity bounds of DRP-ES.


\begin{lemma}
\label{lem:bounded-expectation-tv}
Let \(\mu\) and \(\nu\) be probability measures on the same measurable
space, and let \(g\) be a measurable function satisfying
\[
0\le g\le R
\]
for some finite \(R\ge0\). Then
\[
\left|
\mathbb E_\mu[g]-\mathbb E_\nu[g]
\right|
\le
R\,d_{\mathrm{TV}}(\mu,\nu).
\]
\end{lemma}
\begin{proof}
    See Appendix \ref{app:proof}.
\end{proof}

\begin{lemma}
\label{lem:es-tv-stability}
Suppose that a measurable random variable \(Y\) satisfies $a\le Y\le b$ almost surely under both probability measures \(\mu\) and \(\nu\), where
\(-\infty<a\le b<\infty\). Then
\[
\left|
\operatorname{ES}_{\alpha}(Y\mid\mu)
-
\operatorname{ES}_{\alpha}(Y\mid\nu)
\right|
\le
\frac{b-a}{1-\alpha}
d_{\mathrm{TV}}(\mu,\nu).
\]
\end{lemma}
\begin{proof}
    See Appendix \ref{app:proof}.
\end{proof}

\begin{lemma}
\label{lem:mixture-tv-contraction}
Under Assumption~\ref{ass:bounded-aggregated-loss}, for all \(\mathbf p,\mathbf q\in\Delta_M\), 
we have \[d_{\mathrm{TV}}(P_t^{\mathbf q},P_t^{\mathbf p})\le \frac{1}{2}||\mathbf q-\mathbf p||_1.\]
Hence, we have
\[
|\mathrm{ES}_\alpha \left( Y \mid P_t^{\mathbf q} \right)-\mathrm{ES}_\alpha \left( Y \mid P_t^{\mathbf p} \right)|
\le
\frac{R}{2(1-\alpha)}
||\mathbf q-\mathbf p||_1,
\]
where $R=\overline y-\underline y$ and $Y$ is a measurable random variable.
\end{lemma}
\begin{proof}
    See Appendix \ref{app:proof}.
\end{proof}

\begin{theorem}
\label{thm:drp-disagreement-sensitivity}
Suppose that
Assumption~\ref{ass:bounded-aggregated-loss} holds. Define 
\[
\rho(D)
:=
\min\{
\delta_0+\delta_1D,\log M
\},
\qquad D\ge0.
\]
Then the following statement holds: 
for every \(0\le D_1\le D_2\), let $\rho_i:=\rho(D_i).$ If $\rho_2>0$, 
we have
\[
0
\le
\Psi_{t,\alpha}(D_2, \mathbf X)-\Psi_{t,\alpha}(D_1, \mathbf X)\le
\frac{\overline y-\underline y}{1-\alpha}
\sqrt{
\frac{\delta_1(D_2-D_1)}{2}
},
\]
where $\overline y$ and $\underline y$ are constants defined in Assumption~\ref{ass:bounded-aggregated-loss}. Moreover, for a fixed $D\ge 0$ and for any $0<\alpha_1\le\alpha_2<1$, we have
\[
0\le
\Psi_{t,\alpha_2}(D,\mathbf X)-\Psi_{t,\alpha_1}(D,\mathbf X)
\le
(\overline y-\underline y)
\frac{\alpha_2-\alpha_1}{1-\alpha_1}.
\]

\end{theorem}
\begin{proof}
    See Appendix \ref{app:proof}.
\end{proof}

\subsection{Properties of the DRP-ES}
Since the DRP-ES measures the risk of a random loss vector through an aggregation function, its  properties depend on the choice of the aggregation function.
The following proposition establishes the properties of the DRP-ES when the ambiguity set of model weights is fixed.
\begin{proposition}
\label{prop:properties-of-the-DRP-ES}
Fix $t>0$, $\alpha\in(0,1)$, and $D\geq 0$.
For the general aggregation function $\Lambda_\theta$ and the DRP-ES $\Psi_{t,\alpha}$,
the following statements hold.

\begin{enumerate}
\item[(i)] If $\Lambda_\theta$ satisfies (L0),
then 
\[
\Psi_{t,\alpha}(D,c\mathbf 1_d)=c,
\qquad c\in\mathbb R,
\]
where $\mathbf 1_d=(1,\ldots,1)^\top$.
\item[(ii)] If $\Lambda_\theta$ satisfies (L1),
then $\Psi_{t,\alpha}$ is monotone:
\[
\mathbf X\leq\mathbf Y
\quad\Longrightarrow\quad
\Psi_{t,\alpha}(D,\mathbf X)
\leq
\Psi_{t,\alpha}(D,\mathbf Y).
\]

\item[(iii)] If $\Lambda_\theta$ satisfies (L2),
then $\Psi_{t,\alpha}$ is cash invariant:
\[
\Psi_{t,\alpha}(D,\mathbf X+c\mathbf 1_d)
=
\Psi_{t,\alpha}(D,\mathbf X)+c,
\qquad c\in\mathbb R,
\]
where $\mathbf 1_d=(1,\ldots,1)^\top$.
\item[(iv)] If $\Lambda_\theta$ satisfies (L3),
then $\Psi_{t,\alpha}$ is convex:
\[
\begin{aligned}
\Psi_{t,\alpha}
\left(
D,\lambda\mathbf X+(1-\lambda)\mathbf Y
\right)
\leq
\lambda\Psi_{t,\alpha}(D,\mathbf X)
+
(1-\lambda)\Psi_{t,\alpha}(D,\mathbf Y),
\end{aligned}
\]
where $\lambda\in[0,1]$.

\item[(v)] If $\Lambda_\theta$ satisfies (L4),
then $\Psi_{t,\alpha}$ is positively homogeneous:
\[
\Psi_{t,\alpha}(D,\lambda\mathbf X)
=
\lambda\Psi_{t,\alpha}(D,\mathbf X),
\qquad \lambda\geq0.
\]

\item[(vi)] If $\Lambda_\theta$ satisfies (L5),
then $\Psi_{t,\alpha}$ is subadditive:
\[
\Psi_{t,\alpha}(D,\mathbf X+\mathbf Y)
\leq
\Psi_{t,\alpha}(D,\mathbf X)
+
\Psi_{t,\alpha}(D,\mathbf Y).
\]
\end{enumerate}
\end{proposition}
\begin{proof}
    See Appendix \ref{app:proof}.
\end{proof}
With the exponential aggregation function \eqref{eq:normalized-log-sum-exp-aggregation} which satisfies conditions (L0)-(L3), the DRP-ES is a normalized convex monetary risk measure.
\begin{remark}
In the one-dimensional case ($d=1$), let
$L\in\mathcal X^1$ be a scalar-valued loss random
variable. Then the exponential aggregation function reduces to the linear weighted aggregation function:
\begin{align*}
\Psi_{t,\alpha}(D,L)=\sup_{\mathbf q\in \mathcal{U}_t(D)}\operatorname{ES}_{\alpha}\left(L\mid P_t^{\mathbf q}\right).
\end{align*}
With fixed $D$, the scalar-input DRP-ES $\Psi_{t,\alpha}(D,L)$ is a coherent risk measure.
\end{remark}

\section{Empirical Experiments}
\label{sec:Numerical Experiments}

We conduct empirical experiments on the U.S. stock market, the Chinese A-share market, and the South Korean stock market. We group the stocks in each market into seven sectors, characterized by distinct market capitalizations and investment styles. The seven sectors are Financials, Consumer Discretionary, Consumer Staples, Information Technology, Upstream Cyclicals, Midstream Cyclicals and  Downstream Cyclicals based on the CITIC Tier-1 Industry Classification.

\begin{table}[htbp]
\caption{Sector Classification}\label{tab:sector-class} 
	\resizebox{\textwidth}{!}{
\begin{tabular}{cccc}
\hline
Financials                                                                 & Consumer Discretionary                                                      & Consumer Staples                                                                   & Information Technology     \\ \hline
Banking                                                                   & Automobiles                                                                 & \begin{tabular}[c]{@{}c@{}}Agriculture, Forestry, \\ Fishing, Hunting\end{tabular} & Electronic                 \\
Capital Markets                                                           & Household Durables                                                          & \begin{tabular}[c]{@{}c@{}}Textiles, Apparel, \\ Luxury Goods\end{tabular}         & Telecommunication Services \\
\begin{tabular}[c]{@{}c@{}}Diversified Financial \\ Services\end{tabular} & \begin{tabular}[c]{@{}c@{}}Beverages, Food\\ Staples Retailing\end{tabular} & Pharmaceuticals                                                                    & Technology Hardware        \\
                                                                          &                                                                             & Retailing                                                                          & Media, Entertainment       \\
                                                                          &                                                                             & Consumer
                                                                          Services                                                             &                            \\
                                                                          &                                                                             & Light Industry                                                                     &                            \\ \hline\hline
Upstream Cyclicals                                                        & Midstream Cyclicals                                                         & Downstream Cyclicals                                                               &                            \\ \hline
\begin{tabular}[c]{@{}c@{}}Oil, Gas, \\ Consumable Fuels\end{tabular}     & Chemicals                                                                   & Construction                                                                       &                            \\
Metal, Mining                                                             & Steel                                                             & Real Estate                                                                        &                            \\
                                                                          & Construction Materials                                                      & Utilities                                                                          &                            \\
                                                                          & Machinery                                                                   & Transportation                                                                     &                            \\
                                                                          &Electrical Equipment, 
                                                                          Alternative Electricity                                                      &                                                                                    &                            \\
                                                                          & Aerospace, Defense                                                          &                                                                                    &                            \\ \hline
\end{tabular}
}
\end{table}

We consider two sample periods: January 1, 2019 to August 12, 2026, and January 1, 2012 to August 12, 2019. The reason we choose these two periods is that both of them include extreme market conditions: the Chinese stock market turbulence (2015), the COVID-19 pandemic (2020), geopolitical conflicts (2022), tariff events (2025), the geopolitical crises in the Middle East (2026), and the market crash in the South Korean stock market (2026). We use one year of data as the training set, the following quarter as the validation set to tune the model parameters by maximizing the Sharpe ratio, and the subsequent quarter as the test set.

We construct the following five probability models as inputs of the DRP-ES. 
\begin{enumerate}[(i)] 
\item The historical simulation (HS) that  generates losses directly from the historical loss observations in the training period without assuming a parametric distribution.
		\item  The GARCH model (GARCH) in \cite{bollerslev1986generalized}. 
\item  The extreme value theory model (EVT) in \cite{mcneil1997estimating}.
        \item  The GARCH-EVT model (GARCH-EVT) in \cite{mcneil2000estimation}.
        \item The filtered historical simulation (FHS) in \cite{barone1999var}. 

	\end{enumerate} 
To compare strategies that combine risk model diversification with asset diversification against those that rely solely on asset diversification, we use each candidate model separately as the input  to construct single-scenario expected shortfalls, which are compared with the DRP-ES.
We generate $N=10000$ loss scenarios from each model.  
The parameters $\delta_0$ and $\delta_1$ in \eqref{eq:general_ambiguity_set}, together with the parameters of each candidate model, are optimized using the validation set by maximizing the Sharpe ratio and are then fixed over the subsequent test period. 

We develop a trading strategy based on the DRP-ES. Within each sector, we first compute the individual DRP-ES for every stock:
\begin{align}\label{eq:single-stock-DRP}
\Psi_{t,\alpha}\!\left(D_t,X_i\right)=\sup_{\mathbf q\in\mathcal U_t(D_t)}
\operatorname{ES}_{\alpha}
\left(X_i \mid P_t^{\mathbf q}
\right),
\end{align}
where $X_i$ is the loss of stock $i$, and then we rank the stocks from the lowest to the highest risk. We select \(M_{\text{stock}}\) stocks with the smallest DRP-ES and construct a portfolio in each sector. If there are fewer than \(M_{\text{stock}}\) stocks in a sector, we select all stocks in that sector. Since there are 7 sectors, the total number of selected stocks is no greater than \(7M_{\text{stock}}\). For these stocks, we generate the loss scenarios and aggregate the losses in each scenario using the exponential aggregation function \eqref{eq:normalized-log-sum-exp-aggregation}. Within each portfolio, all selected stocks are assigned equal weights.
We set \(\theta\) in \eqref{eq:normalized-log-sum-exp-aggregation} equal to the median of the standard deviations of the losses of the stocks in the portfolio over the training period.
Then, we apply the same DRP-ES to the aggregated loss and obtain the risk of the portfolio $j$:
\begin{align}\label{eq:aggregate-stock-DRP}
    \Psi_j
    =
    \sup_{\mathbf q\in\mathcal U_t(D_t)}
    \operatorname{ES}_{\alpha}
    \left(
        F_{\theta,\boldsymbol \pi}(\mathbf X_j)
        \mid P_t^{\mathbf q}
    \right), \qquad j=1,2,\cdots,7.
\end{align} After obtaining the risk \(\Psi_j\) for portfolio \(j\), we allocate capital across sectors in inverse proportion to their estimated risks:
$$ w_j = \frac{\frac{1}{\Psi_j}}{\frac{1}{\Psi_1}+\frac{1}{\Psi_2} +\cdots+\frac{1}{\Psi_7} }, $$
where $w_j$ is the weight for the portfolio $j$.
In practice, stocks always have a positive probability of incurring losses. Hence, we only consider the case where $\Psi_j>0$ for all $j=1,2,\ldots,7$.

In the experiment, we adjust stock prices for stock splits and cash dividends, and construct the stock pool by selecting firms with large market capitalizations in each stock market to ensure sufficient stock liquidity.
For both the DRP strategy and the single-probability-model strategies, we hold 21 stocks at each rebalancing date with 3 stocks selected from each sector. After obtaining the weights of all stocks on the previous trading day, we establish the positions at the close of the next trading day.
As for the trading strategies based on the single-scenario expected shortfall, the steps are similar to those of the DRP-ES strategy. We fix the risk-free rates at \(1.7\%\) for the Chinese stock market, \(4.3\%\) for the South Korean stock market, and \(5\%\) for the U.S. stock market. For all trading strategies, we include the actual transaction costs of each stock market and adjust them according to the corresponding historical periods in the backtesting. All transaction costs are calculated as a proportion of the corresponding traded value.

Tables \ref{tab:ES95-2020-2026-all-3} and \ref{tab:ES95-2013-2019-all-3} show the results of different strategies with $\alpha=0.95$. We can observe that in different periods and stock markets, the strategies based on the DRP-ES exhibit lower annualized volatility and maximum drawdown, while achieving higher Sharpe ratios than the benchmarks.  
This performance can be explained by the following reasons: first, the candidate models generate loss distributions that capture different features of market risk, such as time-varying volatility, extreme losses, dependence, and recent historical behavior. Hence, the DRP strategy provides a more comprehensive assessment of portfolio risk and enables more effective risk estimation. Second, the exponential aggregation function places greater emphasis on stocks with larger potential risks without ignoring the contribution of each stock to the overall portfolio risk. These mechanisms help reduce large losses, volatility, and drawdowns of the portfolios while preserving profitability. 

When comparing the DRP strategy with strategies based on the single-scenario  expected shortfall, we observe that the DRP strategy exhibits more robust performance across different markets and time periods. Specifically, the ranking of the Sharpe ratio of the DRP strategy  consistently remains in the upper-middle range among all strategies, with rankings between first and third across different settings in Tables \ref{tab:ES95-2020-2026-all-3} and \ref{tab:ES95-2013-2019-all-3}. In contrast, the performances of the strategies based on the single-scenario  expected shortfall vary more significantly across different situations. 
Regarding the performances of the models across different time periods, during the period from 2013 to 2019, the HS strategy performs well in the Chinese and the South Korean stock markets, ranking first and third among all strategies, respectively. However, during the period from 2020 to 2026, the HS strategy performs worse and ranks only fifth or sixth in terms of the Sharpe ratio among all strategies across the three stock markets. Regarding model performance across different stock markets,
from 2013 to 2019, the EVT strategy ranks second in terms of the Sharpe ratio in the Korean stock market, but only sixth in both the Chinese and U.S. stock markets, outperforming only the benchmarks. In contrast, the GARCH strategy ranks second in both the Chinese and U.S. stock markets, but only sixth in the Korean stock market, again outperforming only the benchmark.
Between 2020 and 2026, the GARCH-EVT strategy and the GARCH strategy also exhibit unstable performances across different markets.
The DRP strategy is more stable because it combines risk information from multiple models rather than relying on a single risk model. Different risk models capture different aspects of market risk, and the DRP framework combines these risk perceptions and provides a more balanced estimate of the downside risk. As a result, its performance is less affected by changes in market conditions and remains more consistent across different markets and periods. These results demonstrate that combining risk model diversification with asset diversification yields better performance than relying solely on asset diversification.

\begin{table}[!h]
\caption{Basic statistics of the strategies in the three stock markets from April 1, 2020 to August 12, 2026 ($\alpha=0.95$). The win rate is the proportion of trading days on which the strategy achieves a positive net return. The ``Rank'' column contains the ranking of the Sharpe ratios of different strategies.} 
\resizebox{\textwidth}{!}{
\begin{tabular}{ccccccc}
\hline\hline
\multicolumn{7}{c}{Chinese stock market}                                                                    \\ \hline
           & Annualized Return & Annualized Volatility & Maximum Drawdown & Sharpe Ratio & Win Rate  & Rank \\ \hline
\textbf{DRP}        & 14.1200$\%$       & 13.0139$\%$           & 12.2735$\%$      & 0.9506       & 51.81$\%$ & \textbf{1}    \\ \hline
HS         & 11.3197$\%$       & 13.0125$\%$           & 14.3231$\%$      & 0.7597       & 52.39$\%$ & 5    \\ \hline
GARCH      & 11.8725$\%$       & 12.8531$\%$           & 12.9101$\%$      & 0.8061       & 52.33$\%$ & 3    \\ \hline
GARCH-EVT  & 11.2279$\%$       & 13.0965$\%$           & 12.9064$\%$      & 0.7493       & 51.74$\%$ & 6    \\ \hline
EVT      & 13.4803$\%$       & 13.0971$\%$           & 11.6144$\%$      & 0.9025       & 51.29$\%$ & 2    \\ \hline
FHS        & 11.9648$\%$       & 13.0475$\%$           & 12.5648$\%$      & 0.8023       & 51.36$\%$ & 4    \\ \hline
CSI 300    & 3.8592$\%$        & 18.0675$\%$           & 45.6025$\%$      & 0.2065       & 50.58$\%$ & 7    \\ \hline\hline
\multicolumn{6}{c}{South Korean stock market}                                                              &      \\ \hline
           & Annualized Return & Annualized Volatility & Maximum Drawdown & Sharpe Ratio & Win Rate  & Rank \\ \hline
\textbf{DRP}        & 21.5700$\%$       & 13.1769$\%$           & 13.3408$\%$      & 1.2295       & 57.33$\%$ & \textbf{1}    \\ \hline
HS         & 20.1128$\%$       & 13.2851$\%$           & 17.3795$\%$      & 1.1294       & 56.43$\%$ & 6    \\ \hline
GARCH      & 21.1314$\%$       & 13.2436$\%$           & 15.3310$\%$      & 1.1966       & 57.71$\%$ & 2    \\ \hline
GARCH-EVT  & 20.9034$\%$       & 13.1924$\%$           & 13.8032$\%$      & 1.1865       & 57.01$\%$ & 3    \\ \hline
EVT      & 20.7130$\%$       & 13.4874$\%$           & 14.5904$\%$      & 1.1517       & 55.79$\%$ & 5    \\ \hline
FHS        & 20.7266$\%$       & 13.4117$\%$           & 17.3087$\%$      & 1.1583       & 56.37$\%$ & 4    \\ \hline
KOSPI      & 23.0743$\%$       & 26.6026$\%$           & 38.6304$\%$      & 0.7554       & 55.60$\%$ & 7    \\ \hline\hline
\multicolumn{7}{c}{U.S. stock market}                                                                         \\ \hline
           & Annualized Return & Annualized Volatility & Maximum Drawdown & Sharpe Ratio & Win Rate  & Rank \\ \hline
\textbf{DRP}        & 15.9037$\%$       & 12.3954$\%$           & 18.1458$\%$      & 0.8593       & 55.03$\%$ & \textbf{3}    \\ \hline
HS         & 15.3279$\%$       & 12.5879$\%$           & 19.2701$\%$      & 0.8085       & 54.34$\%$ & 6    \\ \hline
GARCH      & 15.1238$\%$       & 12.2701$\%$           & 17.0881$\%$      & 0.8118       & 55.15$\%$ & 5    \\ \hline
GARCH-EVT  & 16.2145$\%$       & 12.4202$\%$           & 17.3478$\%$      & 0.8795       & 55.90$\%$ & 1    \\ \hline
EVT      & 15.9292$\%$       & 12.5484$\%$           & 19.0337$\%$      & 0.8521       & 54.09$\%$ & 4    \\ \hline
FHS        & 16.1793$\%$       & 12.5389$\%$           & 18.7836$\%$      & 0.8699       & 55.15$\%$ & 2    \\ \hline
S$\&$P 500 & 18.8245$\%$       & 17.4615$\%$           & 25.4257$\%$      & 0.7958       & 54.59$\%$ & 7    \\ \hline
\end{tabular}
}
\label{tab:ES95-2020-2026-all-3}
\end{table}

\begin{table}[!h]
\caption{Basic statistics of the strategies in the three stock markets from April 1, 2013 to August 12, 2019 ($\alpha=0.95$). The win rate is the proportion of trading days on which the strategy achieves a positive net return. The ``Rank" column contains the ranking of the Sharpe ratios of different strategies.} 
\resizebox{\textwidth}{!}{
\begin{tabular}{ccccccc}
\hline\hline
\multicolumn{7}{c}{Chinese stock market}                                                                    \\ \hline
           & Annualized Return & Annualized Volatility & Maximum Drawdown & Sharpe Ratio & Win Rate  & Rank \\ \hline
\textbf{DRP}        & 17.4923$\%$       & 20.0006$\%$           & 38.7882$\%$      & 0.8225       & 53.83$\%$ & \textbf{3}    \\ \hline
HS         & 19.9882$\%$       & 17.4386$\%$           & 24.5286$\%$      & 1.0362       & 52.22$\%$ & 1    \\ \hline
GARCH      & 19.0359$\%$       & 19.9583$\%$           & 44.4565$\%$      & 0.8894       & 53.90$\%$ & 2    \\ \hline
GARCH-EVT  & 16.4176$\%$       & 19.2253$\%$           & 44.6522$\%$      & 0.7999       & 51.70$\%$ & 5    \\ \hline
EVT      & 15.9832$\%$       & 20.5548$\%$           & 39.8960$\%$      & 0.7430       & 53.57$\%$ & 6    \\ \hline
FHS        & 17.3962$\%$       & 20.2010$\%$           & 39.4795$\%$      & 0.8125       & 54.02$\%$ & 4    \\ \hline
CSI 300    & 6.3755$\%$        & 23.5849$\%$           & 46.6960$\%$      & 0.3093       & 51.83$\%$ & 7    \\ \hline\hline
\multicolumn{6}{c}{South Korean stock market}                                                              &      \\ \hline
           & Annualized Return & Annualized Volatility & Maximum Drawdown & Sharpe Ratio & Win Rate  & Rank \\ \hline
\textbf{DRP}        & 6.0535$\%$        & 8.9980$\%$            & 13.7979$\%$      & 0.2303       & 53.51$\%$ & \textbf{1}    \\ \hline
HS         & 4.9864$\%$        & 9.2671$\%$            & 14.5543$\%$      & 0.1171       & 52.87$\%$ & 3    \\ \hline
GARCH      & 3.6972$\%$        & 8.8409$\%$            & 14.0461$\%$      & -0.0213      & 54.09$\%$ & 6    \\ \hline
GARCH-EVT  & 3.7396$\%$        & 9.0576$\%$            & 14.7883$\%$      & -0.0141      & 53.70$\%$ & 5    \\ \hline
EVT      & 5.0056$\%$        & 8.9894$\%$            & 13.8838$\%$      & 0.1199       & 53.32$\%$ & 2    \\ \hline
FHS        & 4.5006$\%$        & 9.0482$\%$            & 14.2357$\%$      & 0.0665       & 54.09$\%$ & 4    \\ \hline
KOSPI      & -0.4966$\%$       & 11.9425$\%$           & 26.4984$\%$      & -0.3344      & 51.79$\%$ & 7    \\ \hline\hline
\multicolumn{7}{c}{U.S. stock market}                                                                         \\ \hline
           & Annualized Return & Annualized Volatility & Maximum Drawdown & Sharpe Ratio & Win Rate  & Rank \\ \hline
\textbf{DRP}        & 11.8535$\%$       & 10.4254$\%$           & 11.8271$\%$      & 0.6588       & 55.42$\%$ & \textbf{3}    \\ \hline
HS         & 11.8417$\%$       & 10.6895$\%$           & 13.7586$\%$      & 0.6442       & 54.36$\%$ & 4    \\ \hline
GARCH      & 12.2664$\%$       & 10.5626$\%$           & 13.2876$\%$      & 0.6865       & 55.92$\%$ & 2    \\ \hline
GARCH-EVT  & 11.7815$\%$       & 10.7182$\%$           & 14.0638$\%$      & 0.6377       & 55.11$\%$ & 5    \\ \hline
EVT      & 10.7365$\%$       & 10.5049$\%$           & 12.6715$\%$      & 0.5590       & 54.67$\%$ & 6    \\ \hline
FHS        & 12.5888$\%$       & 10.5273$\%$           & 12.3858$\%$      & 0.7157       & 55.73$\%$ & 1    \\ \hline
S$\&$P 500 & 10.0181$\%$       & 12.9732$\%$           & 19.7795$\%$      & 0.4249       & 53.80$\%$ & 7    \\ \hline
\end{tabular}
}
\label{tab:ES95-2013-2019-all-3}
\end{table}

\begin{figure}[!h]
    \centering

    \begin{subfigure}[b]{0.8\textwidth}
        \centering
        \includegraphics[width=\textwidth]{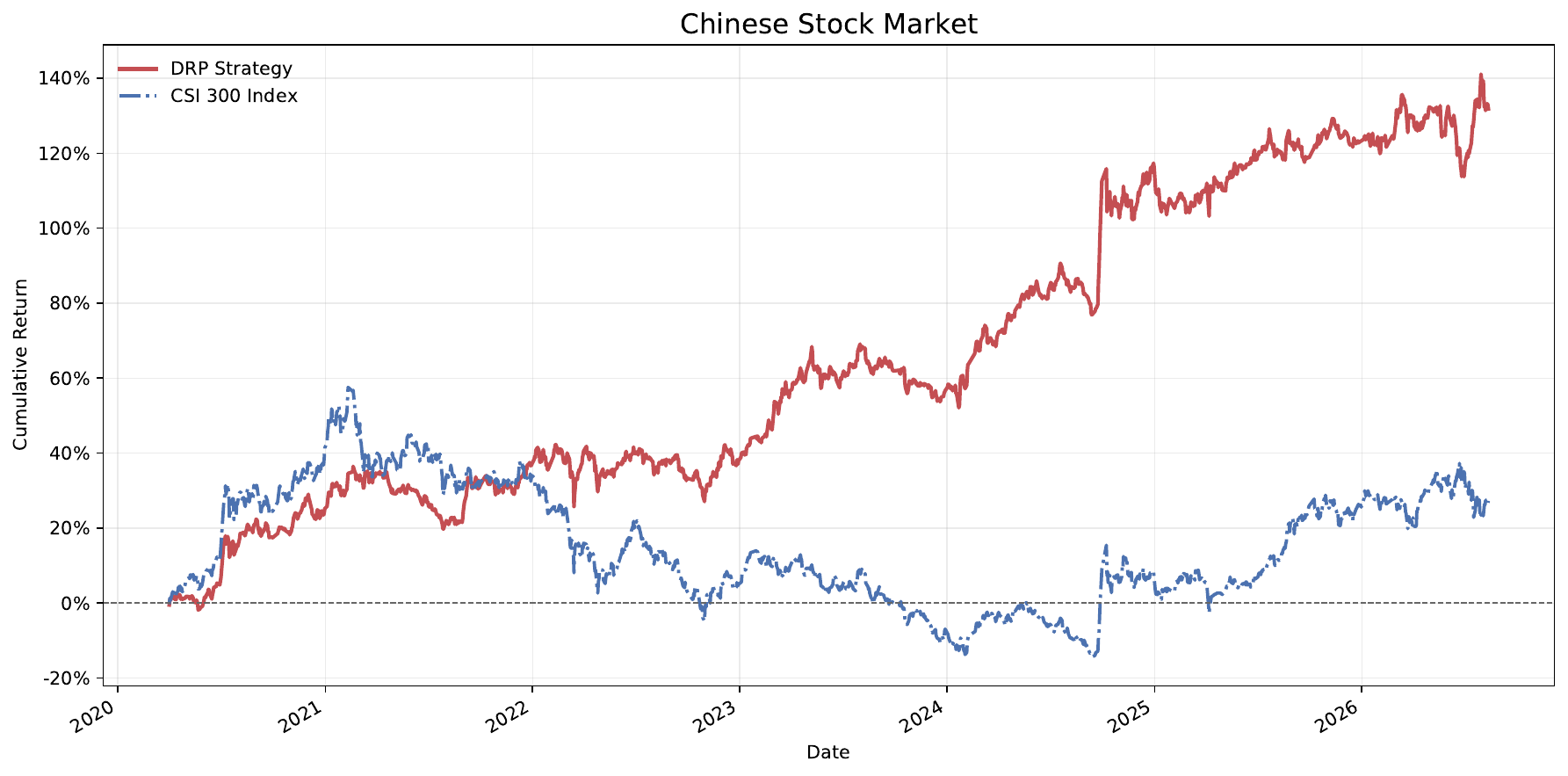}
        \caption{Comparison between the DRP strategy and the benchmark from April 1, 2020 to August 12, 2026.}
    \end{subfigure}

    \vspace{0.3cm}

    \begin{subfigure}[b]{0.8\textwidth}
        \centering
        \includegraphics[width=\textwidth]{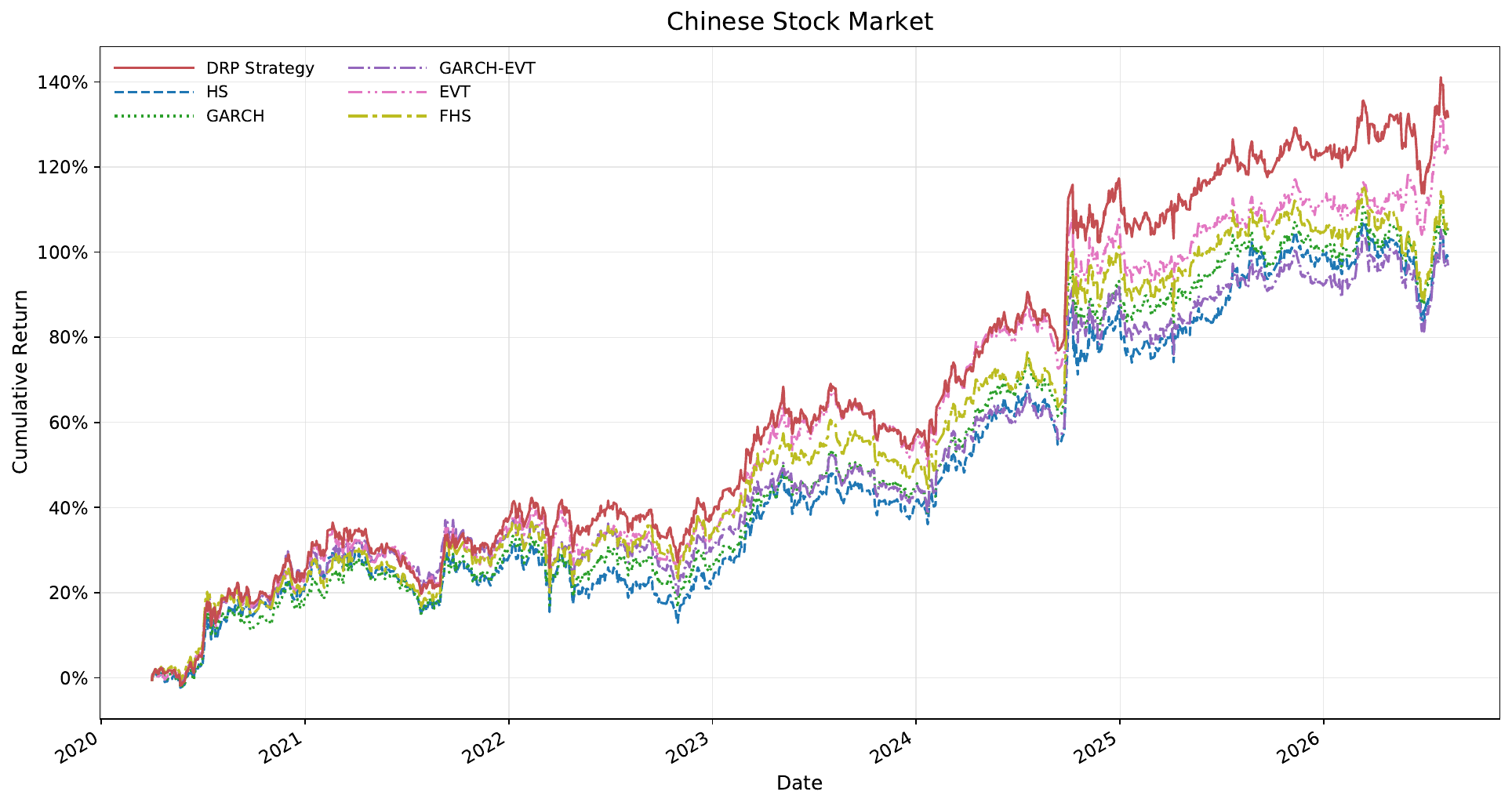}
        \caption{Comparison between the DRP strategy and the single-model strategies  from April 1, 2020 to August 12, 2026.}
        \label{fig:second}
    \end{subfigure}

    \caption{DRP strategy in the Chinese stock market  from April 1, 2020 to August 12, 2026 ($\alpha=0.95$).}
    \label{fig:ES975-2020-2026-China-25}
\end{figure}

\begin{figure}[!h]
    \centering

    \begin{subfigure}[b]{0.8\textwidth}
        \centering
        \includegraphics[width=\textwidth]{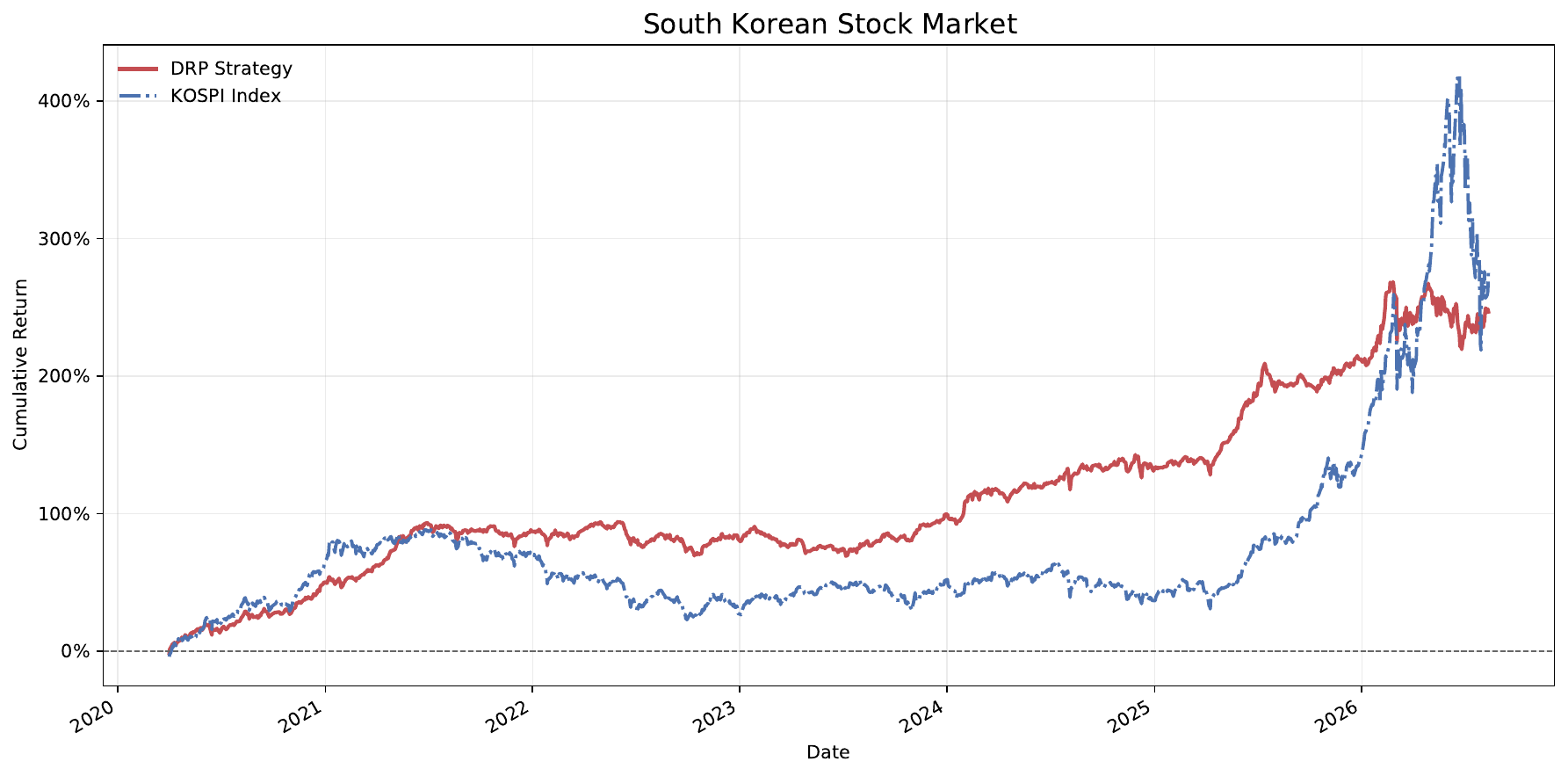}
        \caption{Comparison between the DRP strategy and the benchmark  from April 1, 2020 to August 12, 2026.}
    \end{subfigure}

    \vspace{0.3cm}

    \begin{subfigure}[b]{0.8\textwidth}
        \centering
        \includegraphics[width=\textwidth]{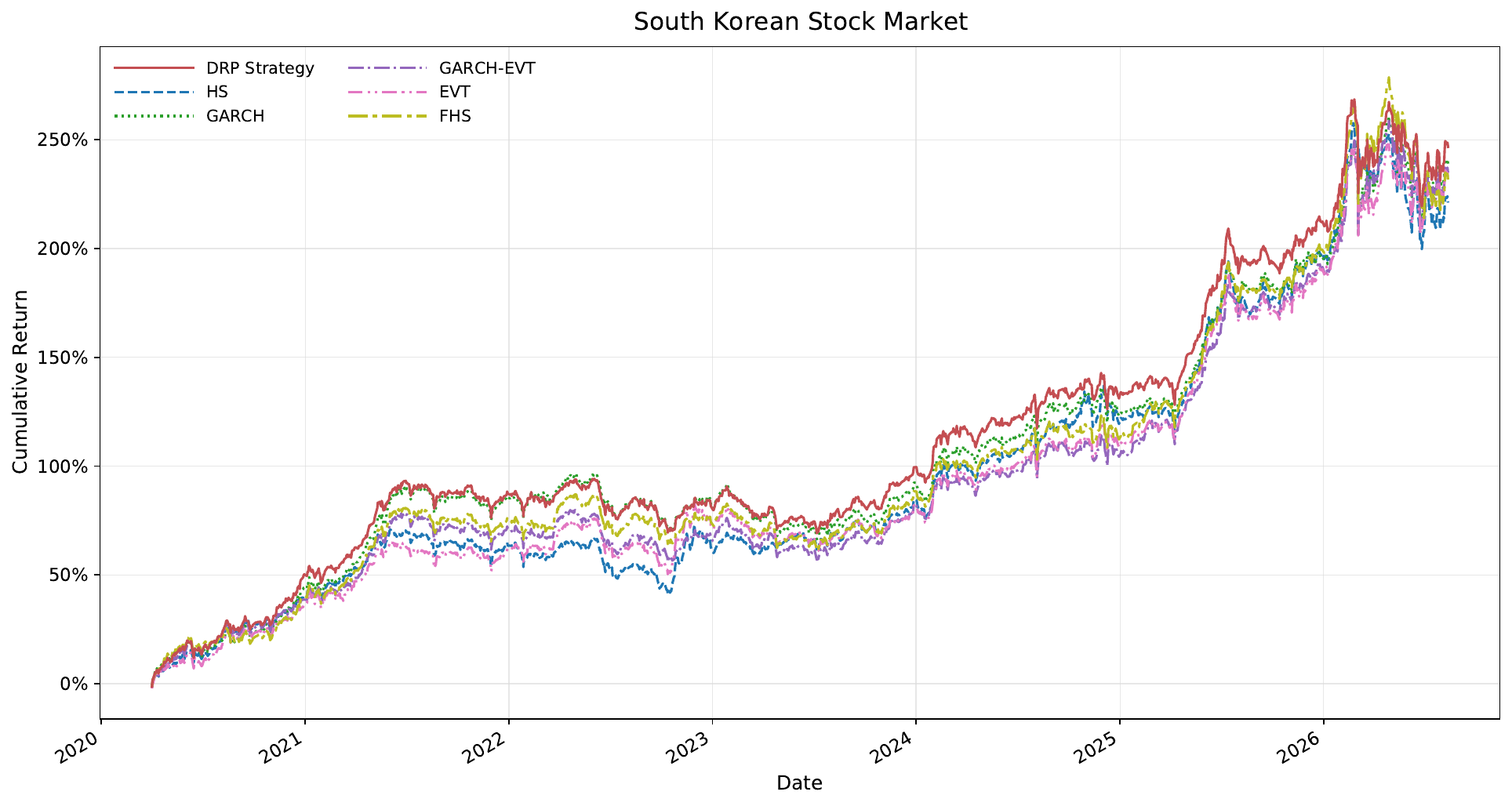}
        \caption{Comparison between the DRP strategy and the single-model strategies  from April 1, 2020 to August 12, 2026.}
    \end{subfigure}

    \caption{DRP strategy in the South Korean stock market  from April 1, 2020 to August 12, 2026 ($\alpha=0.95$).}
    \label{fig:ES975-2020-2026-Korea-25}
\end{figure}

\begin{figure}[!h]
    \centering

    \begin{subfigure}[b]{0.8\textwidth}
        \centering
        \includegraphics[width=\textwidth]{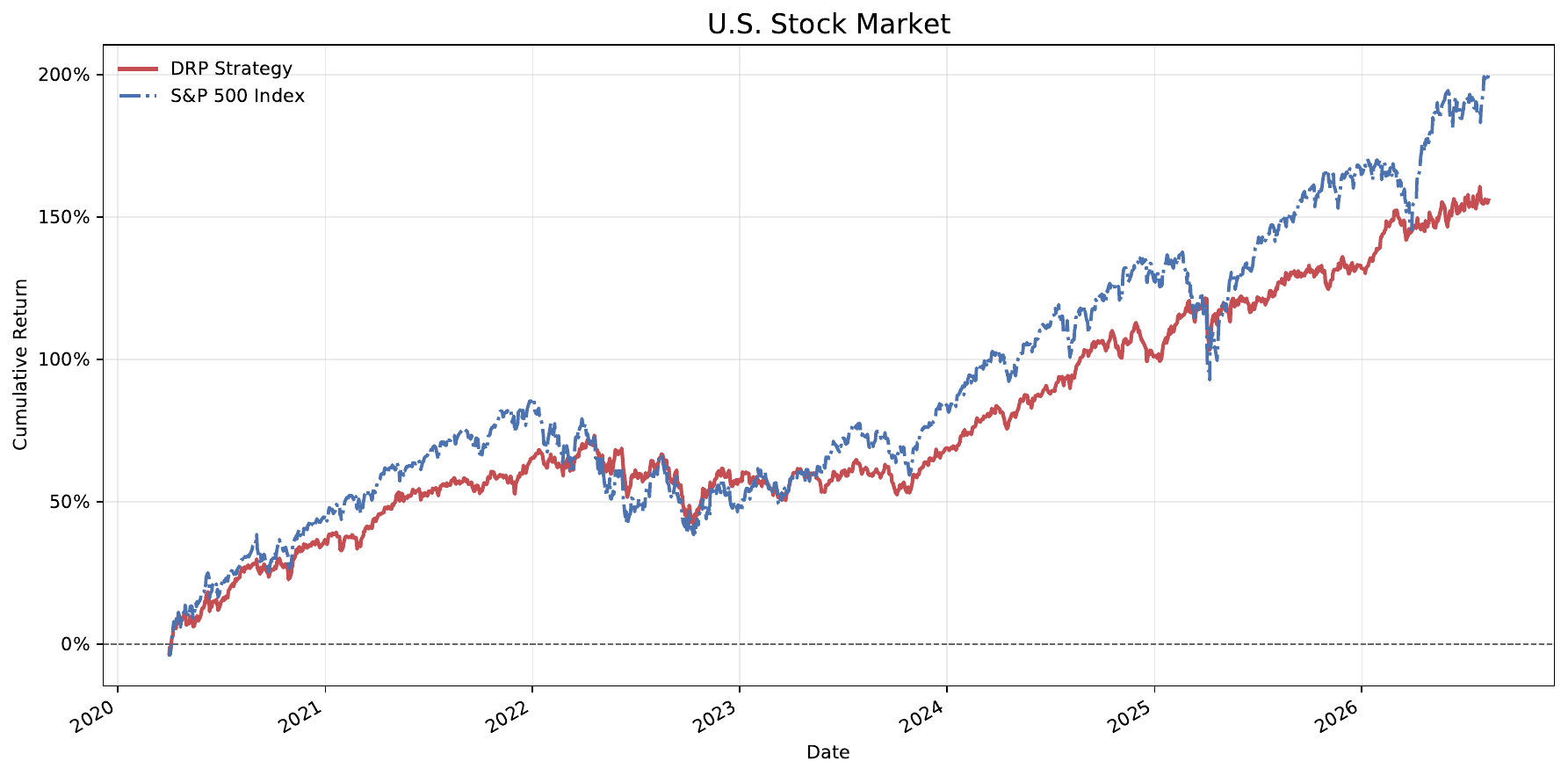}
        \caption{Comparison between the DRP strategy and the benchmark  from April 1, 2020 to August 12, 2026.}
    \end{subfigure}

    \vspace{0.3cm}

    \begin{subfigure}[b]{0.8\textwidth}
        \centering
        \includegraphics[width=\textwidth]{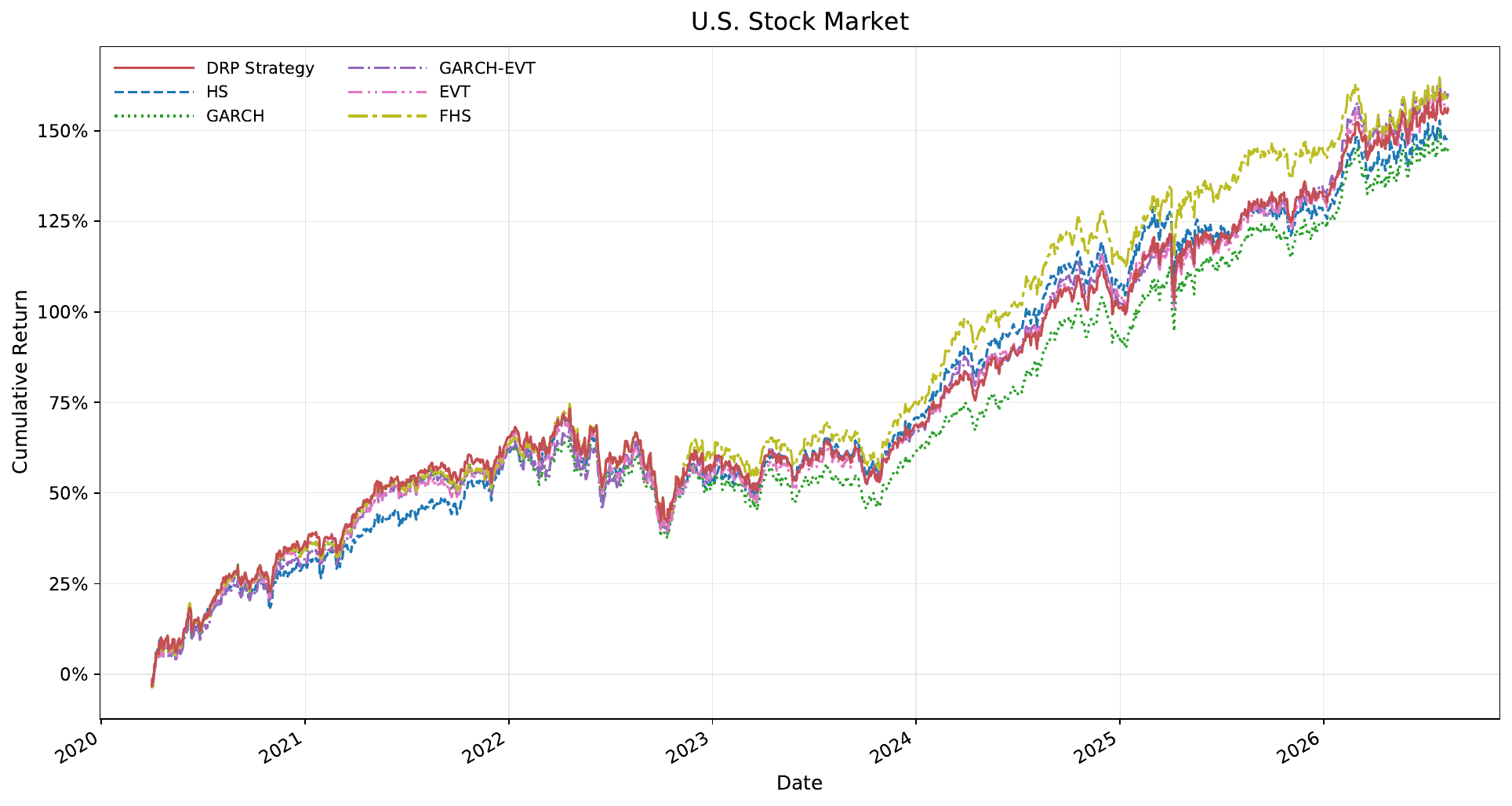}
        \caption{Comparison between the DRP strategy and the single-model strategies  from April 1, 2020 to August 12, 2026.}
    \end{subfigure}

    \caption{DRP strategy in the U.S. stock market  from April 1, 2020 to August 12, 2026 ($\alpha=0.95$).}
    \label{fig:ES975-2020-2026-US-25}
\end{figure}

Figures \ref{fig:ES975-2020-2026-China-25}--\ref{fig:ES975-2020-2026-US-25} compare the performances of different strategies and benchmarks across the three stock markets over the period from April 1, 2020 to August 12, 2026. We can observe that when the market was rising, the cumulative return of the DRP strategy also increased. When the market declined, the DRP strategy experienced smaller losses than the benchmark. For example, the South Korean stock market experienced a substantial decline in 2026, but the drawdown of the DRP strategy was noticeably smaller than that of the benchmark during this period. These results indicate that the DRP strategy is effective in controlling downside risk. In addition, we find that the cumulative return curve of the DRP strategy generally lies in the upper-middle range among those of the single-model strategies, indicating stable performance of the DRP strategy.

For the sensitivity analysis, we change the confidence level of expected shortfall from \(\alpha=0.95\) to \(\alpha=0.975\) and \(\alpha=0.9\) respectively, while keeping all other settings unchanged. Table \ref{tab:ES90-975-sharp-ratio-3} reports the strategy performance when the expected shortfall confidence level is set to \(\alpha=0.975\) and \(\alpha=0.9\). We observe that the DRP strategy continues to outperform the benchmarks and exhibit more stable Sharpe ratio rankings than the single-model expected shortfall strategies across different markets and periods. Specifically, its Sharpe ratio consistently ranks between first and fourth among all the strategies, indicating relatively stable performance under the alternative confidence levels. 

\begin{table}[!h]
\caption{Sharpe ratios of the strategies in different stock markets with different $\alpha$. The ``Rank" column contains the ranking of the Sharpe ratios of the strategies  across different models. The benchmarks for the Chinese stock market, the South Korean stock market, and the U.S. stock market are CSI 300, KOSPI, and S\&P 500 respectively. } 
\resizebox{\textwidth}{!}{
\begin{tabular}{ccccccccccccc}
\hline
\multicolumn{13}{c}{April 1, 2020 to August 12, 2026}                                                                                                                                                                                                                                         \\ \hline
\multicolumn{1}{c|}{}          & \multicolumn{6}{c||}{$\alpha$=0.975}                                                                                            & \multicolumn{6}{c}{$\alpha$=0.9}                                                                                          \\ \cline{2-13} 
\multicolumn{1}{c|}{}          & \multicolumn{2}{c|}{Chinese stock market} & \multicolumn{2}{c|}{South Korean stock market} & \multicolumn{2}{c||}{U.S. stock market}    & \multicolumn{2}{c|}{Chinese stock market} & \multicolumn{2}{c|}{South Korean stock market} & \multicolumn{2}{c}{U.S. stock market} \\ \hline
\multicolumn{1}{c|}{Strategy}  & Sharpe ratio  & \multicolumn{1}{c|}{Rank}  & Sharpe ratio  & \multicolumn{1}{c|}{Rank} & Sharpe ratio & \multicolumn{1}{c||}{Rank} & Sharpe ratio  & \multicolumn{1}{c|}{Rank}  & Sharpe ratio  & \multicolumn{1}{c|}{Rank} & Sharpe ratio          & Rank         \\ \hline
\multicolumn{1}{c|}{\textbf{DRP}}       & 0.9028       & \textbf{1}                          & 1.2383       & \textbf{2}                         & 0.8330      & \multicolumn{1}{c||}{\textbf{3}}    & 0.8607       & \textbf{2}                          & 1.2364       & \textbf{1}                         & 0.8452               & \textbf{4}            \\ \hline
\multicolumn{1}{c|}{HS}        & 0.8024       & 3                          & 1.0969       & 5                         & 0.7885      & \multicolumn{1}{c||}{7}    & 0.7332       & 6                          & 1.1385       & 5                         & 0.8103               & 5            \\ \hline
\multicolumn{1}{c|}{GARCH}     & 0.7953       & 4                          & 1.2137       & 3                         & 0.8679      & \multicolumn{1}{c||}{2}    & 0.7837       & 3                          & 1.2061       & 3                         & 0.8645               & 3            \\ \hline
\multicolumn{1}{c|}{GARCH-EVT} & 0.7704       & 5                          & 1.1310       & 4                         & 0.8328      & \multicolumn{1}{c||}{4}    & 0.7650       & 5                          & 1.1528       & 4                         & 0.8998               & 2            \\ \hline
\multicolumn{1}{c|}{EVT}     & 0.8184       & 2                          & 1.2460       & 1                         & 0.8236      & \multicolumn{1}{c||}{5}    & 0.9151       & 1                          & 1.0572       & 6                         & 0.8008               & 6            \\ \hline
\multicolumn{1}{c|}{FHS}       & 0.7561       & 6                          & 1.0268       & 6                         & 0.9454      & \multicolumn{1}{c||}{1}    & 0.7680       & 4                          & 1.2309       & 2                         & 0.9238               & 1            \\ \hline
\multicolumn{1}{c|}{Benchmark} & 0.2065       & 7                          & 0.7554       & 7                         & 0.7958      & \multicolumn{1}{c||}{6}    & 0.2065       & 7                          & 0.7554       & 7                         & 0.7958               & 7            \\ \hline
\multicolumn{13}{c}{April 1, 2013 to August 12, 2019}                                                                                                                                                                                                                                         \\ \hline
\multicolumn{1}{c|}{}          & \multicolumn{6}{c||}{$\alpha$=0.975}                                                                                            & \multicolumn{6}{c}{$\alpha$=0.9}                                                                                          \\ \cline{2-13} 
\multicolumn{1}{c|}{}          & \multicolumn{2}{c|}{Chinese stock market} & \multicolumn{2}{c|}{South Korean stock market} & \multicolumn{2}{c||}{U.S. stock market}    & \multicolumn{2}{c|}{Chinese stock market} & \multicolumn{2}{c|}{South Korean stock market} & \multicolumn{2}{c}{U.S. stock market} \\ \hline
\multicolumn{1}{c|}{Strategy}  & Sharpe ratio  & \multicolumn{1}{c|}{Rank}  & Sharpe ratio  & \multicolumn{1}{c|}{Rank} & Sharpe ratio & \multicolumn{1}{c||}{Rank} & Sharpe ratio  & \multicolumn{1}{c|}{Rank}  & Sharpe ratio  & \multicolumn{1}{c|}{Rank} & Sharpe ratio          & Rank         \\ \hline
\multicolumn{1}{c|}{\textbf{DRP}}       & 0.8519       & \textbf{4}                          & 0.3646       & \textbf{1}                         & 0.5889      & \multicolumn{1}{c||}{\textbf{4}}    & 0.8329       & \textbf{3}                          & 0.2734       & \textbf{1}                         & 0.6877               & \textbf{2}            \\ \hline
\multicolumn{1}{c|}{HS}        & 1.1113       & 1                          & 0.1422       & 3                         & 0.6758      & \multicolumn{1}{c||}{2}    & 0.9886       & 1                          & 0.0925       & 3                         & 0.6737               & 4            \\ \hline
\multicolumn{1}{c|}{GARCH}     & 0.8119       & 6                          & -0.0784      & 6                         & 0.6853      & \multicolumn{1}{c||}{1}    & 0.9532       & 2                          & 0.0163       & 5                         & 0.7353               & 1            \\ \hline
\multicolumn{1}{c|}{GARCH-EVT} & 0.8719       & 3                          & 0.1344       & 4                         & 0.6113      & \multicolumn{1}{c||}{3}    & 0.7211       & 5                          & -0.0507      & 6                         & 0.6442               & 5            \\ \hline
\multicolumn{1}{c|}{EVT}     & 0.8220       & 5                          & 0.2669       & 2                         & 0.5455      & \multicolumn{1}{c||}{5}    & 0.7057       & 6                          & 0.1275       & 2                         & 0.5386               & 6            \\ \hline
\multicolumn{1}{c|}{FHS}       & 0.9051       & 2                          & 0.1324       & 5                         & 0.5375      & \multicolumn{1}{c||}{6}    & 0.7479       & 4                          & 0.0677       & 4                         & 0.6755               & 3            \\ \hline
\multicolumn{1}{c|}{Benchmark} & 0.3093       & 7                          & -0.3344      & 7                         & 0.4249      & \multicolumn{1}{c||}{7}    & 0.3093       & 7                          & -0.3344      & 7                         & 0.4249               & 7            \\ \hline
\end{tabular}}
\label{tab:ES90-975-sharp-ratio-3}
\end{table}

Finally, we compare the performance of the strategy under different aggregation functions and different model weights including equal weights, worst-case weights, and the optimized weights used in the DRP approach. Under equal weights, each of the candidate models is assigned the same weight, that is, \(q_m=1/M\) for all \(m\) in \eqref{eq:mixture_probability}. Under worst-case weights, all the weight is assigned to the candidate model with the highest expected shortfall, while the other models receive zero weight. Table \ref{tab:stat-DRP-average-maximum-aggregation-ES975-2020-2026}  shows that the DRP strategy using the exponential aggregation function and optimized weights performs well across the three markets. In terms of the Sharpe ratio, the exponential aggregation function outperforms both the linear weighted aggregation function and the maximum aggregation function in the Chinese and U.S. stock markets. In the South Korean market, its Sharpe ratio is lower only than that of the linear weighted aggregation function. Similarly, the optimized weights outperform the equal weights and the worst-case weights in the Chinese and U.S. stock markets, while the worst-case weights achieve the highest Sharpe ratio in the South Korean market. These results could be explained by the flexibility of the exponential aggregation function and the optimized weights. The exponential aggregation function approaches the maximum aggregation function as \(\theta\) tends to zero and the linear weighted aggregation function as \(\theta\) tends to infinity. Likewise, the optimized weights approach equal weights when model disagreement is small and worst-case weights when model disagreement is sufficiently large. Therefore, the exponential aggregation function and optimized weights can adapt to different risk environments, which  explains the more stable and generally stronger performance of the  DRP strategy.

\begin{table}[!h]
\centering
\caption{Basic statistics of the DRP strategies with different aggregation functions or different weighting methods for the probability models in different stock markets from April 1, 2020 to August 12, 2026 ($\alpha=0.95$). The ``Rank" column contains the ranking of the Sharpe ratios. The win rate is the proportion of trading days on which the strategy achieves a positive net return.} 
\resizebox{\textwidth}{!}{
\begin{tabular}{ccccccc}
\hline
\multicolumn{6}{c}{Chinese stock market}                                                                              \\ \hline
                          Aggregation Function  & Annualized Return & Annualized Volatility & Maximum Drawdown & Sharpe Ratio & Win Rate &Rank  \\ \hline
\textbf{Exponential aggregation function} &  14.1200$\%$       & 13.0139$\%$           & 12.2735$\%$      & 0.9506       & 51.81$\%$    &   \textbf{1}    \\ \hline
Linear weighted aggregation function & 13.5079$\%$       & 12.9160$\%$           & 11.5650$\%$      & 0.9152       & 51.68$\%$  &3 \\ \hline
Maximum aggregation function & 13.7874$\%$       & 12.9569$\%$           & 11.8497$\%$      & 0.9317       & 51.81$\%$     &  2    \\ \hline
CSI 300    & 3.8592$\%$        & 18.0675$\%$           & 45.6025$\%$      & 0.2065       & 50.58$\%$  &  4\\ \hline
\multicolumn{6}{c}{South Korean stock market}                                                                               \\ \hline
                           Aggregation Function & Annualized Return & Annualized Volatility & Maximum Drawdown & Sharpe Ratio & Win Rate &Rank \\ \hline
\textbf{Exponential aggregation function} & 21.5700$\%$       & 13.1769$\%$           & 13.3408$\%$      & 1.2295       & 57.33$\%$  & \textbf{2}\\ \hline
Linear weighted aggregation function & 21.4677$\%$       & 13.0728$\%$           & 13.4343$\%$      & 1.2318       & 57.07$\%$ & 1 \\ \hline
Maximum aggregation function & 21.3878$\%$       & 13.1233$\%$           & 13.7234$\%$      & 1.2225       & 57.01$\%$       &   3 \\ \hline
KOSPI      & 23.0743$\%$       & 26.6026$\%$           & 38.6304$\%$      & 0.7554       & 55.60$\%$ &4 \\ \hline
\multicolumn{6}{c}{U.S. stock market}                                                                                   \\ \hline
                          Aggregation Function  & Annualized Return & Annualized Volatility & Maximum Drawdown & Sharpe Ratio & Win Rate &Rank \\ \hline
\textbf{Exponential aggregation function}   & 15.9037$\%$       & 12.3954$\%$           & 18.1458$\%$      & 0.8593       & 55.03$\%$     &   \textbf{1}   \\ \hline
Linear weighted aggregation function & 15.4438$\%$       & 12.4099$\%$           & 18.3329$\%$      & 0.8264       & 54.97$\%$ &2\\ \hline
Maximum aggregation function & 15.4116$\%$       & 12.4166$\%$           & 18.1907$\%$      &0.8238       & 54.65$\%$     &  3    \\ \hline
S$\&$P 500 & 18.8245$\%$       & 17.4615$\%$           & 25.4257$\%$      & 0.7958       & 54.59$\%$&4 \\ \hline
\multicolumn{6}{c}{Chinese stock market}                                                                              \\ \hline
Probability Model Weights & Annualized Return & Annualized Volatility & Maximum Drawdown & Sharpe Ratio & Win Rate &Rank \\ \hline
\textbf{Optimized weights} & 14.1200$\%$       & 13.0139$\%$           & 12.2735$\%$      & 0.9506       & 51.81$\%$     &  \textbf{1}   \\ \hline
Equal weights & 13.4972$\%$       & 12.9373$\%$           & 12.5182$\%$      & 0.9132       & 51.61$\%$ &2\\ \hline
Worst-case weights & 13.3031$\%$       & 13.0909$\%$           &11.5946$\%$      & 0.8909       & 51.68$\%$      &  3   \\ \hline
CSI 300    & 3.8592$\%$        & 18.0675$\%$           & 45.6025$\%$      & 0.2065       & 50.58$\%$& 4\\ \hline
\multicolumn{6}{c}{South Korean stock market}                                                                               \\ \hline
                          Probability Model Weights  & Annualized Return & Annualized Volatility & Maximum Drawdown & Sharpe Ratio & Win Rate &Rank \\ \hline
\textbf{Optimized weights} & 21.5700$\%$       & 13.1769$\%$           & 13.3408$\%$      & 1.2295       & 57.33$\%$ &\textbf{2}\\ \hline
Equal weights & 21.4259$\%$       & 13.2733$\%$           & 12.8162$\%$      & 1.2126       & 57.33$\%$ &3\\ \hline
Worst-case weights & 22.0324$\%$       & 13.3015$\%$           & 13.8059$\%$      &1.2477       & 57.27$\%$       &  1  \\ \hline
KOSPI      & 23.0743$\%$       & 26.6026$\%$           & 38.6304$\%$      & 0.7554       & 55.60$\%$ &4\\ \hline
\multicolumn{6}{c}{U.S. stock market}                                                                                   \\ \hline
                           Probability Model Weights & Annualized Return & Annualized Volatility & Maximum Drawdown & Sharpe Ratio & Win Rate &Rank \\ \hline
\textbf{Optimized weights}   & 15.9037$\%$       & 12.3954$\%$           & 18.1458$\%$      & 0.8593       & 55.03$\%$     &  \textbf{1}     \\ \hline
Equal weights &15.4409$\%$       & 12.3083$\%$           & 16.9964$\%$      & 0.8320       & 54.34$\%$& 2\\ \hline
Worst-case weights & 14.5054$\%$       & 12.4038$\%$           & 19.4855$\%$      &0.7610       & 54.40$\%$    &  4     \\ \hline
S$\&$P 500 & 18.8245$\%$       & 17.4615$\%$           & 25.4257$\%$      & 0.7958       & 54.59$\%$ & 3\\ \hline
\end{tabular}
}
\label{tab:stat-DRP-average-maximum-aggregation-ES975-2020-2026}
\end{table}

\section{Conclusion}\label{sec:conclusion}

We establish a Diversified-Risk-Perception (DRP) risk measure framework that redefines conventional paradigms by combining asset diversification with risk model diversification. By integrating heterogeneous risk control models as decision-making criteria, we not only expose cognitive biases in market perception across model types but also develop an integrative methodology for risk diversification, fundamentally challenging the traditional optimal model paradigm. As for  asset diversification,  we introduce a flexible exponential aggregation function to evaluate the risk of a portfolio, which places greater emphasis on large individual losses while accounting for the contributions of all assets. As for  risk model diversification, we incorporate different probability models in the risk measure and develop a method for determining their weights based on their perceptions on the risk assessments.
To validate this approach, we evaluate its performance in the U.S., Chinese A-share, and South Korean stock markets over two different periods, which include different extreme market scenarios. Empirical results demonstrate that the strategies based on DRP-ES consistently achieve higher Sharpe ratios and lower maximum drawdowns than the market benchmarks across all three markets and both periods. 
Compared with strategies based on the individual probability models, the DRP strategy exhibits more stable and robust performance across different markets and periods, while the performances of the strategies based on the single-model expected shortfall vary considerably with market conditions. These findings indicate that while single risk models face a high risk of failure across different economic cycles, the risk diversification endows the investment strategy with exceptional environmental adaptability, serving as a profound economic manifestation of anti-fragility. Overall, combining risk model diversification with asset diversification  can improve the consistency of risk assessments and the effectiveness of downside risk management. Future work includes using modern reinforcement learning techniques (e.g., \cite{liu2026noise}) to elicit risk aversion and depict different risk perceptions in the generalized-risk-measure framework. 




\vspace{6mm}
\noindent{\normalsize \bf Acknowledgements} \vskip0.1in\parskip=0mm

\noindent  
The authors would like to thank 
Songzhe Liu for her excellent research assistance on the paper. Yang Liu acknowledges financial support from the National Natural Science Foundation of China (Grant No. 12401624), Guangdong Science and Technology Program (Grant No. 2024QN11X076), Shenzhen Science and Technology Program (Grant Nos. RCBS20231211090814028, JCYJ20250604141203005, 2025TC0010) and The Chinese University of Hong Kong (Shenzhen) University Development Fund (Grant No. UDF01003336) and is partly supported by the Guangdong Provincial Key Laboratory of Mathematical Foundations for Artificial Intelligence (Grant No. 2023B1212010001).

\begin{appendices}

\section{Proofs}\label{app:proof}
\begin{proof}[Proof of Proposition \ref{prop:Limiting-relations-of-log-sum-exp-aggregations}]
Let $m=\max_{1\leq i\leq d}X_i.$ Suppose that 
$X_k=m$. Then
\[
    \pi_k e^{m/\theta}
    \leq
    \sum_{i=1}^d
    \pi_i e^{X_i/\theta}
    \leq
    e^{m/\theta}.
\]
It follows that
\[
    m+\theta\log\pi_k
    \leq
    F_{\theta,\boldsymbol{\pi}}(\mathbf{X})
    \leq
    m.
\]
Letting $\theta\downarrow0$ gives
\[
    F_{\theta,\boldsymbol{\pi}}(\mathbf{X})
    \longrightarrow
    \max_{1\leq i\leq d}X_i.
\]

Using
\[
    e^{X_i/\theta}=
    1+\frac{X_i}{\theta}
    +O(\theta^{-2}),
\]
we have
\[
    \sum_{i=1}^d
    \pi_i e^{X_i/\theta}
    =
    1+
    \frac{1}{\theta}
    \sum_{i=1}^d\pi_i X_i
    +
    O(\theta^{-2}).
\]
Since $\log(1+z)=z+O(z^2)$ as $z\to0$, it follows that
\[
\begin{aligned}
    F_{\theta,\boldsymbol{\pi}}(\mathbf{X})
    &=
    \theta
    \log
    \left[
        1+
        \frac{\boldsymbol{\pi}^\top\mathbf{X}}{\theta}
        +
        O(\theta^{-2})
    \right] =
    \boldsymbol{\pi}^\top\mathbf{X}
    +
    O(\theta^{-1}).
\end{aligned}
\]
Hence,
\[
    \lim_{\theta\to\infty}
    F_{\theta,\boldsymbol{\pi}}(\mathbf{X})
    =
    \boldsymbol{\pi}^\top\mathbf{X}.
\]
\end{proof}

\begin{proof}[Proof of Theorem \ref{thm:limiting_optimal_weights}]
    We first consider the case in which
\(D\downarrow0\) and \(\delta_0\downarrow0\).
Since
\(\mathbf q_{D}^*\in\mathcal U_t(D)\), we have
\[
    \operatorname{KL}
    \left(
        \mathbf q_{D}^*
        \mid
        \boldsymbol{\omega}_0
    \right)
    \leq
    \delta_0+\delta_1D.
\]
By Pinsker's inequality,
\[
    \frac{1}{2}
    ||
        \mathbf q_{D}^*-
        \boldsymbol{\omega}_0
    ||_1
    \leq
    \sqrt{
        \frac{1}{2}
        \operatorname{KL}
        \left(
            \mathbf q_{D}^*
            \mid
            \boldsymbol{\omega}_0
        \right)
    }
    \leq
    \sqrt{
        \frac{\delta_0+\delta_1D}{2}
    }.
\]
Since $\delta_0+\delta_1D\longrightarrow0$
as \(D\downarrow0\) and \(\delta_0\downarrow0\), it follows that
\[
    ||
        \mathbf q_{D}^*-
        \boldsymbol{\omega}_0
    ||_1
    \longrightarrow0.
\]
Hence, we have $\mathbf q_{D}^*\longrightarrow\boldsymbol{\omega}_0.$

We next consider the case \(D\to\infty\). By Lemma \ref{lem:upper-bound-of-KL-radius},
\[
    \operatorname{KL}(\mathbf q\mid\boldsymbol\omega_0)
    \leq \log M,
    \qquad
    \mathbf q\in\Delta_M.
\]
Therefore, once $\delta_0+\delta_1D\geq \log M,$
we have $\mathcal U_t(D)=\Delta_M.$
By the variational representation of the expected shortfall,
\[
    \operatorname{ES}_{\alpha}
    \left(Y\mid P_t^{\mathbf q}\right)
    =
    \min_{z}\left\{
    z+\frac{1}{1-\alpha}\mathbb E_{P_t^{\mathbf q}}\left[(Y-z)_+\right]\right\}.
\]
Since \(z_\alpha^*\) satisfies
\[
    P_{m^*,t}(Y<z_\alpha^*)
    \leq \alpha
    \leq
    P_{m^*,t}(Y\leq z_\alpha^*),
\]
we have
\[
    \operatorname{ES}_{\alpha}
    \left(Y\mid P_{m^*,t}\right)
    =
    z_\alpha^*
    +
    \frac{1}{1-\alpha}
    \mathbb E_{P_{m^*,t}}
    \left[(Y-z_\alpha^*)_+\right].
\]

For any \(\mathbf q\in\Delta_M\),
\[
\begin{aligned}
    \operatorname{ES}_{\alpha}
    \left(Y\mid P_t^{\mathbf q}\right)
    &\leq
    z_\alpha^*+\frac{1}{1-\alpha}\mathbb E_{P_t^{\mathbf q}}\left[(Y-z_\alpha^*)_+\right]\\
    &=
    z_\alpha^*
    +
    \frac{1}{1-\alpha}
    \sum_{m=1}^M q_m
    \mathbb E_{P_{m,t}}
    \left[(Y-z_\alpha^*)_+\right].
\end{aligned}
\]
Note that
\[
    \mathbb E_{P_{m,t}}
    \left[(Y-z_\alpha^*)_+\right]
    <
    \mathbb E_{P_{m^*,t}}
    \left[(Y-z_\alpha^*)_+\right]
\]
for every \(m\neq m^*\). Therefore, for $\mathbf q\ne \mathbf e_{m^*}$, we have
\[
\begin{aligned}
    \operatorname{ES}_{\alpha}
    \left(Y\mid P_t^{\mathbf q}\right)
    &<
    z_\alpha^*
    +
    \frac{1}{1-\alpha}
    \mathbb E_{P_{m^*,t}}
    \left[(Y-z_\alpha^*)_+\right]\\
    &=
    \operatorname{ES}_{\alpha}
    \left(Y\mid P_{m^*,t}\right)=\operatorname{ES}_{\alpha}
    \left(Y\mid P_t^{\mathbf e_{m^*}}\right).
\end{aligned}
\]
Note that \(\mathbf e_{m^*}\in \Delta_M\). Consequently, for $D\geq\frac{(\log M-\delta_0)_+}{\delta_1},$
we have
\[
    \mathbf q_D^*=\mathbf e_{m^*},
\]
and therefore, we have
\[
    \mathbf q_D^*\longrightarrow\mathbf e_{m^*},
    \qquad D\to\infty.
\]
\end{proof}

\begin{proof}[Proof of Lemma \ref{lem:bounded-expectation-tv}]
If \(R=0\), then \(g=0\) and the conclusion follows. Suppose that \(R>0\)
and define \(h:=g/R\). Then \(0\le h\le1\), and the layer-cake
representation gives
\[
h(\omega)=\int_0^1
\mathbf 1_{\{h(\omega)>s\}}\,\d s.
\]
The integrand is nonnegative and measurable. Hence, Tonelli's theorem yields
\[
\mathbb E_\mu[h]
=
\int_0^1\mu(h>s)\,\d s, \qquad
\mathbb E_\nu[h]
=
\int_0^1\nu(h>s)\,\d s.
\]
Consequently,
\[
\begin{aligned}
\left|
\mathbb E_\mu[h]-\mathbb E_\nu[h]
\right|
&=
\left|
\int_0^1
\bigl(
\mu(h>s)-\nu(h>s)
\bigr)\,\d s
\right|\\
&\le
\int_0^1
\left|
\mu(h>s)-\nu(h>s)
\right|\,\d s\\
&\le
\int_0^1
d_{\mathrm{TV}}(\mu,\nu)\,\d s=
d_{\mathrm{TV}}(\mu,\nu).
\end{aligned}
\]
Multiplying both sides by \(R\) proves the result.
\end{proof}

\begin{proof}[Proof of Lemma \ref{lem:es-tv-stability}]
By the variational representation of expected shortfall,
\[
\operatorname{ES}_{\alpha}(Y\mid\mu)
=
\min_{z\in\mathbb R}
\left\{
z+
\frac1{1-\alpha}
\mathbb E_\mu[(Y-z)_+]
\right\}.
\]
Define
\[
\phi_\mu(z)
:=
z+
\frac1{1-\alpha}
\mathbb E_\mu[(Y-z)_+].
\]

We first show that the minimization is restricted to \(z\in[a,b]\).
If \(z<a\), then \(Y\ge a>z\) almost surely and hence
\[
(Y-z)_+=Y-z.
\]
It follows that
\[
\phi_\mu(z)
=
\frac{\mathbb E_\mu[Y]}{1-\alpha}
-
\frac{\alpha}{1-\alpha}z.
\]
Since \(-\alpha/(1-\alpha)<0\), the function \(\phi_\mu\) is strictly
decreasing on \((-\infty,a]\).

If \(z>b\), then \((Y-z)_+=0\) almost surely, and hence
\[
\phi_\mu(z)=z.
\]
Consequently, we have
\[
\operatorname{ES}_{\alpha}(Y\mid\mu)
=
\min_{z\in[a,b]}\phi_\mu(z).
\]
Since $|(Y-z)_+-(Y-z')_+|\le |z-z'|,$ we have
\begin{align*}
|\phi_\mu(z) - \phi_\mu(z')| &\leq |z - z'| + \frac{1}{1-\alpha} \mathbb{E}_\mu \left[ |(Y - z)_+ - (Y - z')_+| \right] \\
&\leq |z - z'| + \frac{1}{1-\alpha} |z - z'| \\
&= \left(1 + \frac{1}{1-\alpha}\right) |z - z'|.
\end{align*}
Hence, \(\phi_\mu\) is continuous, and \(\phi_\mu\) attains its minimum on \([a,b]\).
For every \(z\in[a,b]\), we have
\[
0\le (Y-z)_+\le b-a.
\]
Lemma~\ref{lem:bounded-expectation-tv} therefore gives
\[
\left|
\mathbb E_\mu[(Y-z)_+]
-
\mathbb E_\nu[(Y-z)_+]
\right|
\le
(b-a)d_{\mathrm{TV}}(\mu,\nu).
\]
Let \(z_\nu\in[a,b]\) be a minimizer of \(\phi_\nu\). Then
\[
\begin{aligned}
\operatorname{ES}_{\alpha}(Y\mid\mu)
&=
\min_{z\in[a,b]}\phi_\mu(z)\\
&\le
\phi_\mu(z_\nu)\\
&=
\phi_\nu(z_\nu)
+
\frac1{1-\alpha}
\left(
\mathbb E_\mu[(Y-z_\nu)_+]
-
\mathbb E_\nu[(Y-z_\nu)_+]
\right)\\
&\le
\operatorname{ES}_{\alpha}(Y\mid\nu)
+
\frac{b-a}{1-\alpha}
d_{\mathrm{TV}}(\mu,\nu).
\end{aligned}
\]
Interchanging the roles of \(\mu\) and \(\nu\) gives
\[
\operatorname{ES}_{\alpha}(Y\mid\nu)
\le
\operatorname{ES}_{\alpha}(Y\mid\mu)
+
\frac{b-a}{1-\alpha}
d_{\mathrm{TV}}(\mu,\nu).
\]
Combining the two inequalities proves the claim.
\end{proof}

\begin{proof}[Proof of Lemma \ref{lem:mixture-tv-contraction}]
As for $\mathbf p=(p_1,p_2,\cdots ,p_M)$ and $\mathbf q=(q_1,q_2,\cdots,q_M)$, let $\delta_m=q_m-p_m.$
Since \(\mathbf p,\mathbf q\in\Delta_M\), we have
\[
\sum_{m=1}^M\delta_m=0.
\]
Therefore,
\[
\sum_{\delta_m>0}\delta_m
=
-\sum_{\delta_m<0}\delta_m
=
\frac12\sum_{m=1}^M|\delta_m|
=
\frac{1}{2}||\mathbf q-\mathbf p||_1.
\]
For any event \(A\), by definition, we have
\[
P_t^{\mathbf q}(A)-P_t^{\mathbf p}(A)=
\sum_{m=1}^M
\delta_mP_{m,t}(A).
\]
Since \(0\le P_{m,t}(A)\le1\),
\[
\sum_{m=1}^M
\delta_mP_{m,t}(A)
\le
\sum_{\delta_m>0}\delta_m
=
\frac{1}{2}||\mathbf q-\mathbf p||_1,
\]
and
\[
\sum_{m=1}^M
\delta_mP_{m,t}(A)
\ge
\sum_{\delta_m<0}\delta_m
=
-\frac{1}{2}||\mathbf q-\mathbf p||_1.
\]
Hence, for any $A$, we have
\[
|P_t^{\mathbf q}(A)-P_t^{\mathbf p}(A)|
\le
\frac{1}{2}||\mathbf q-\mathbf p||_1,
\]
and
\[
d_{\mathrm{TV}}(P_t^{\mathbf q},P_t^{\mathbf p})\le
\frac{1}{2}||\mathbf q-\mathbf p||_1.
\]
The second statement follows from
Lemma~\ref{lem:es-tv-stability}.
\end{proof}

\begin{proof}[Proof of Theorem \ref{thm:drp-disagreement-sensitivity}]

If \(D_1\le D_2\), then \(\rho_1\le\rho_2\), and  $\mathcal U_t(D_1)\subseteq \mathcal U_t(D_2).$
Taking the supremum over the set $\mathcal U_t$ gives
\[
\Psi_{t,\alpha}(D_1, \mathbf X)\le\Psi_{t,\alpha}(D_2, \mathbf X).
\]
Suppose that
\(\rho_2>0\), and take any \(\mathbf q\in\Delta_M\) satisfying
\[
\operatorname{KL}(\mathbf q\mid\boldsymbol\omega_0)\le\rho_2.
\]
Set $s:=\frac{\rho_1}{\rho_2}\le 1$,
and
\[
\mathbf q^{(s)}:=
s\mathbf q+(1-s)\boldsymbol\omega_0.
\]
Let \(f(x)=x\log x\), with \(f(0)=0\). Since \(f\) is convex, we have
\[
\begin{aligned}
\operatorname{KL}(\mathbf q^{(s)}\mid\boldsymbol\omega_0)
&=
\sum_{m=1}^M
\omega_{0,m}
f\left(
s\frac{q_m}{\omega_{0,m}}+(1-s)
\right)\\
&\le s
\sum_{m=1}^M
\omega_{0,m}
f\left(
\frac{q_m}{\omega_{0,m}}
\right)
+(1-s)
\sum_{m=1}^M
\omega_{0,m}f(1)\\
&=s\operatorname{KL}(\mathbf q\mid\boldsymbol\omega_0)\\
&\le
s\rho_2=
\rho_1.
\end{aligned}
\]
Furthermore, we have
\[
\mathbf q-\mathbf q^{(s)}
=
(1-s)(\mathbf q-\boldsymbol\omega_0),
\]
and therefore, by Pinsker's inequality, we have
\[
\begin{aligned}
\frac12||\mathbf q-\mathbf q^{(s)}||_1&=\frac{1-s}{2}||\mathbf q-\boldsymbol\omega_0||_1\\
&\le
(1-s)
\sqrt{
\frac12
\operatorname{KL}(\mathbf q\mid\boldsymbol\omega_0)
}\\
&\le
(1-s)\sqrt{\frac{\rho_2}{2}}=
\frac{\rho_2-\rho_1}{\sqrt{2\rho_2}}.
\end{aligned}
\]

By Lemma~\ref{lem:mixture-tv-contraction},
\[
\begin{aligned}
\mathrm{ES}_\alpha(\Lambda_{\theta}(\mathbf X)\mid P_t^{\mathbf q})
&\le \mathrm{ES}_\alpha(\Lambda_{\theta}(\mathbf X)\mid P_t^{\mathbf q^{(s)}})
+
\frac{\overline y-\underline y}{2(1-\alpha)}
||\mathbf q-\mathbf q^{(s)}||_1\\
&\le
\Psi_{t,\alpha}(D_1, \mathbf X)
+
\frac{\overline y-\underline y}{1-\alpha}
\frac{\rho_2-\rho_1}{\sqrt{2\rho_2}}.
\end{aligned}
\]
This inequality holds for every \(\mathbf q\) in the larger KL ball. Taking the
supremum over all such \(\mathbf q\) yields
\[
\Psi_{t,\alpha}(D_2, \mathbf X)
\le
\Psi_{t,\alpha}(D_1, \mathbf X)
+
\frac{\overline y-\underline y}{1-\alpha}
\frac{\rho_2-\rho_1}{\sqrt{2\rho_2}}.
\]
Combining this with monotonicity proves
\[
0
\le
\Psi_{t,\alpha}(D_2, \mathbf X)-\Psi_{t,\alpha}(D_1, \mathbf X)
\le
\frac{\overline y-\underline y}{1-\alpha}
\frac{\rho_2-\rho_1}{\sqrt{2\rho_2}}.
\]

Since \(0\le\rho_2-\rho_1\le\rho_2\) and $\rho_2>0$, we have
\[
\frac{\rho_2-\rho_1}{\sqrt{\rho_2}}
\le
\sqrt{\rho_2-\rho_1}.
\]
Moreover, the mapping \(x\mapsto\min\{x,\log M\}\) is
\(1\)-Lipschitz, and hence
\[
\rho_2-\rho_1
\le
\delta_1(D_2-D_1).
\]
Therefore,
\[
0
\le
\Psi_{t,\alpha}(D_2, \mathbf X)-\Psi_{t,\alpha}(D_1, \mathbf X)
\le
\frac{\overline y-\underline y}{1-\alpha}
\sqrt{
\frac{\delta_1(D_2-D_1)}{2}
}.
\]

Fix $\mathbf q\in\mathcal U_t(D)$. 
Let
\[
Q_{\mathbf q}(u)=\operatorname{VaR}_u(\Lambda_{\theta}(\mathbf X)\mid P_t^{\mathbf q}),
\qquad
e_i(\mathbf q)=\operatorname{ES}_{\alpha_i}(\Lambda_{\theta}(\mathbf X)\mid P_t^{\mathbf q}),
\quad i=1,2.
\]
By the quantile representation of the expected shortfall,
\begin{align}\label{eq:quantile-representation-of-the-expected-shortfall}
(1-\alpha_1)e_1(\mathbf q)
=\int_{\alpha_1}^{\alpha_2}Q_{\mathbf q}(u)\,\d u  +\int_{\alpha_2}^{1}Q_{\mathbf q}(u)\,\d u=\int_{\alpha_1}^{\alpha_2}Q_{\mathbf q}(u)\,\d u
  +(1-\alpha_2)e_2(\mathbf q).
\end{align}
Rearranging Equation \eqref{eq:quantile-representation-of-the-expected-shortfall} gives
\[
e_2(\mathbf q)-e_1(\mathbf q)
=
\frac{1}{1-\alpha_1}
\int_{\alpha_1}^{\alpha_2}
\bigl(e_2(\mathbf q)-Q_{\mathbf q}(u)\bigr)\,\d u.
\]
For $u\in[\alpha_1,\alpha_2]$, the monotonicity of $Q_{\mathbf q}$
implies
\[
Q_{\mathbf q}(u)
\le
\frac{1}{1-\alpha_2}
\int_{\alpha_2}^{1}Q_{\mathbf q}(v)\,\d v
=e_2(\mathbf q).
\]
By Assumption \ref{ass:bounded-aggregated-loss}, we have $Q_{\mathbf q}(u)\ge\underline y$ and
$e_2(\mathbf q)\le\overline y$. Consequently,
\begin{align}\label{eq:bound-of-the-difference-ES}
0\le e_2(\mathbf q)-e_1(\mathbf q)
\le
(\overline y-\underline y)
\frac{\alpha_2-\alpha_1}{1-\alpha_1}.
\end{align}
This estimate holds uniformly over $\mathbf q\in\mathcal U_t(D)$.
Taking the supremum  in \eqref{eq:bound-of-the-difference-ES}
proves the second result.
\end{proof}

\begin{proof}[Proof of Proposition \ref{prop:properties-of-the-DRP-ES}]
\textit{(i)}
By (L0), for every $c\in\mathbb R$, we have $\Lambda_\theta(c\mathbf 1_d)=c$.
Hence, 
\[
\begin{aligned}
&\Psi_{t,\alpha}(D,c\mathbf 1_d)=
\sup_{\mathbf q\in\mathcal U_t(D)}
\operatorname{ES}_{\alpha}
\left(
\Lambda_\theta(c\mathbf 1_d)
\mid P_t^{\mathbf q}
\right)=
\sup_{\mathbf q\in\mathcal U_t(D)}
\operatorname{ES}_{\alpha}
\left(c\mid P_t^{\mathbf q}\right)=c.
\end{aligned}
\]

\textit{(ii)}
Suppose that $\mathbf X\leq\mathbf Y$
componentwise.
It follows from (L1) that $\Lambda_{\theta}(\mathbf X)\leq\Lambda_\theta(\mathbf Y).$
The monotonicity of the expected shortfall implies
that, for every $\mathbf q\in\mathcal U_t(D)$, we have
\[
\begin{aligned}
&\operatorname{ES}_{\alpha}
\left(
\Lambda_{\theta}(\mathbf X)
\mid P_t^{\mathbf q}
\right)\leq
\operatorname{ES}_{\alpha}
\left(
\Lambda_\theta(\mathbf Y)
\mid P_t^{\mathbf q}
\right).
\end{aligned}
\]
Taking the supremum over
$\mathbf q\in\mathcal U_t(D)$ yields
\[
\Psi_{t,\alpha}(D,\mathbf X)
\leq
\Psi_{t,\alpha}(D,\mathbf Y).
\]

\textit{(iii)}
By (L2), for every $c\in\mathbb R$, we have $\Lambda_\theta(\mathbf X+c\mathbf 1_d)=\Lambda_{\theta}(\mathbf X)+c.$
Using the cash invariance of expected shortfall,
we obtain
\[
\begin{aligned}
\Psi_{t,\alpha}(D,\mathbf X+c\mathbf 1_d)&=
\sup_{\mathbf q\in\mathcal U_t(D)}
\operatorname{ES}_{\alpha}
\left(
\Lambda_{\theta}(\mathbf X)+c
\mid P_t^{\mathbf q}
\right)\\
&=\sup_{\mathbf q\in\mathcal U_t(D)}
\left[
\operatorname{ES}_{\alpha}
\left(
\Lambda_{\theta}(\mathbf X)
\mid P_t^{\mathbf q}
\right)+c
\right]=
\Psi_{t,\alpha}(D,\mathbf X)+c.
\end{aligned}
\]

\textit{(iv)}
Let $\lambda\in[0,1]$.
By (L3), we have
\[
\begin{aligned}
&\Lambda_\theta
\left(
\lambda\mathbf X+(1-\lambda)\mathbf Y
\right)\leq
\lambda\Lambda_{\theta}(\mathbf X)
+
(1-\lambda)\Lambda_\theta(\mathbf Y).
\end{aligned}
\]
Using the monotonicity and convexity of
expected shortfall, for every
$\mathbf q\in\mathcal U_t(D)$, we obtain
\[
\begin{aligned}
&\operatorname{ES}_{\alpha}
\left(
\Lambda_\theta
\left(\lambda\mathbf X+(1-\lambda)\mathbf Y\right)
\mid P_t^{\mathbf q}
\right)\\
&\leq
\operatorname{ES}_{\alpha}
\left(
\lambda\Lambda_{\theta}(\mathbf X)
+(1-\lambda)\Lambda_\theta(\mathbf Y)
\mid P_t^{\mathbf q}
\right)\\
&\leq
\lambda\operatorname{ES}_{\alpha}
\left(
\Lambda_{\theta}(\mathbf X)
\mid P_t^{\mathbf q}
\right)+
(1-\lambda)\operatorname{ES}_{\alpha}
\left(
\Lambda_\theta(\mathbf Y)
\mid P_t^{\mathbf q}
\right).
\end{aligned}
\]
Taking the supremum over
$\mathbf q\in\mathcal U_t(D)$ and using
the subadditivity of the supremum gives
\[
\begin{aligned}
\Psi_{t,\alpha}
\left(
D,\lambda\mathbf X+(1-\lambda)\mathbf Y
\right)&\leq
\lambda
\sup_{\mathbf q\in\mathcal U_t(D)}
\operatorname{ES}_{\alpha}
\left(
\Lambda_{\theta}(\mathbf X)
\mid P_t^{\mathbf q}
\right)+
(1-\lambda)
\sup_{\mathbf q\in\mathcal U_t(D)}
\operatorname{ES}_{\alpha}
\left(
\Lambda_\theta(\mathbf Y)
\mid P_t^{\mathbf q}
\right)\\
&=
\lambda\Psi_{t,\alpha}(D,\mathbf X)
+
(1-\lambda)\Psi_{t,\alpha}(D,\mathbf Y).
\end{aligned}
\]

\textit{(v)}
Let $\lambda\geq0$.
(L4) implies that $\Lambda_\theta(\lambda\mathbf X)=\lambda\Lambda_{\theta}(\mathbf X).$
The positive homogeneity of expected shortfall
gives
\[
\begin{aligned}
\Psi_{t,\alpha}(D,\lambda\mathbf X)=
\sup_{\mathbf q\in\mathcal U_t(D)}
\operatorname{ES}_{\alpha}
\left(
\lambda\Lambda_{\theta}(\mathbf X)
\mid P_t^{\mathbf q}
\right)=
\sup_{\mathbf q\in\mathcal U_t(D)}
\lambda\operatorname{ES}_{\alpha}
\left(
\Lambda_{\theta}(\mathbf X)
\mid P_t^{\mathbf q}
\right)=
\lambda\Psi_{t,\alpha}(D,\mathbf X).
\end{aligned}
\]

\textit{(vi)}
By (L5), we have
\[
\Lambda_\theta(\mathbf X+\mathbf Y)
\leq
\Lambda_{\theta}(\mathbf X)
+
\Lambda_\theta(\mathbf Y).
\]
Using the monotonicity and subadditivity of
expected shortfall, for every
$\mathbf q\in\mathcal U_t(D)$, we have
\[
\begin{aligned}
\operatorname{ES}_{\alpha}
\left(
\Lambda_\theta(\mathbf X+\mathbf Y)
\mid P_t^{\mathbf q}
\right)&\leq
\operatorname{ES}_{\alpha}
\left(
\Lambda_{\theta}(\mathbf X)+\Lambda_\theta(\mathbf Y)
\mid P_t^{\mathbf q}
\right)\\
&\leq
\operatorname{ES}_{\alpha}
\left(
\Lambda_{\theta}(\mathbf X)
\mid P_t^{\mathbf q}
\right)+
\operatorname{ES}_{\alpha}
\left(
\Lambda_\theta(\mathbf Y)
\mid P_t^{\mathbf q}
\right).
\end{aligned}
\]
Taking the supremum over
$\mathbf q\in\mathcal U_t(D)$ yields
\[
\begin{aligned}
\Psi_{t,\alpha}(D,\mathbf X+\mathbf Y)&\leq
\sup_{\mathbf q\in\mathcal U_t(D)}
\operatorname{ES}_{\alpha}
\left(
\Lambda_{\theta}(\mathbf X)
\mid P_t^{\mathbf q}
\right)+
\sup_{\mathbf q\in\mathcal U_t(D)}
\operatorname{ES}_{\alpha}
\left(
\Lambda_\theta(\mathbf Y)
\mid P_t^{\mathbf q}
\right)\\
&=
\Psi_{t,\alpha}(D,\mathbf X)
+
\Psi_{t,\alpha}(D,\mathbf Y).
\end{aligned}
\]

Combining these results proves all the statements.
\end{proof}

\end{appendices}

\bibliographystyle{chicagoa}
	\bibliography{name}

\begin{thebibliography}{10}
	
	
	\bibitem[\protect\citeauthoryear{Artzner et al.}{Artzner et al.}{1999}]{ADEH99}{Artzner, P., Delbaen, F., Eber, J.-M. and Heath, D.} (1999). Coherent measures of risk. \emph{Mathematical Finance}, \textbf{9}(3), 203--228.
	
	
	
	
	
	
	
	
	
	
	
	
	%
	
	
	
	
	
	
	
	
	\bibitem[\protect\citeauthoryear{Delbaen}{2002}]{D02}Delbaen, F. (2002).   Coherent risk measures on general probability spaces. In \emph{Advances in Finance and Stochastics} (pp. 1--37). Springer, Berlin, Heidelberg.
	
	
	
	
	
	
	
	
	
	
	
	
	
	
	\bibitem[\protect\citeauthoryear{Fadina et al.}{2024}]{FLW21} Fadina, T., Liu, Y. and Wang, R. (2024). A framework for measures of risk under uncertainty.  \emph{Finance and Stochastics}, \textbf{28}(2), 363--390.
	
	\bibitem[\protect\citeauthoryear{F\"ollmer and Schied}{F\"ollmer and Schied}{2016}]{FS16} F\"ollmer, H. and Schied, A. (2016). \emph{Stochastic Finance. An Introduction in Discrete Time}.  {Walter de Gruyter, Berlin}, Fourth Edition.
	
	
	
	{\color{red}\bibitem[\protect\citeauthoryear{Tahar et al.}{2014}]{TL14}	Tahar, Imen Ben and L\'{e}pinette, Emmanuel. (2014). Vector-valued coherent risk measure processes. \emph{International Journal of Theoretical and Applied Finance}, \textbf{17}(2). }
	
	
	
	
	
	
	
	
	
	
	
	
	
	
	
	\bibitem[\protect\citeauthoryear{Henry et al.}{2011}]{HEG11} Ekeland, I., Galichon, A., and Henry, M. (2011). Comonotonic measures of multivariate risks. \emph{Mathematical Finance}.\textbf{22}(1), 109--132.
	
	
	
	
	
	
	\bibitem[\protect\citeauthoryear{Jouini et al.}{2004}]{JMN04} Jouini, E., Meddeb, M. and Touzi, N. (2004). Vector-valued coherent risk measures. \emph{Finance and stochastics}, \textbf{8}(4), 531--552.
	
	
	
	
	
	
	%
	
	
	
	
	
	
	
	
	
	
	
	
	
	
	
	
	
	
	
	
	
	
	
	
	
	
	\bibitem[\protect\citeauthoryear{Shen et al.}{2019}]{SSWW19} 
	Shen, J., Shen, Y., Wang, B. and Wang, R. (2019). Distributional compatibility for change of measures. \emph{Finance and Stochastics}, \textbf{23}(3), 761--794.
	
	
	
	
	\bibitem[\protect\citeauthoryear{Wang and Wei}{2020}]{WW20}
	Wang, R. and Wei, Y. (2020).
	Risk functionals with convex level sets. \emph{Mathematical Finance}, \textbf{30}(4), 1337--1367.
	
	
	\bibitem[\protect\citeauthoryear{Wang et al.}{Wang et al.}{2020}]{WWW20} Wang, R., Wei, Y. and Willmot, G. E. (2020). Characterization, robustness and aggregation of signed Choquet integrals. \emph{Mathematics of Operations Research}, \textbf{45}(3), 993--1015.
	
	
	
	\bibitem[\protect\citeauthoryear{Wang and Ziegel}{2021}]{WZ18} Wang, R. and Ziegel, J. F. (2021). Scenario-based risk evaluation. \emph{Finance and Stochastics}, \textbf{25}, 725--756. 
	
	\bibitem[\protect\citeauthoryear{Wang and Zitikis}{2021}]{WZ21} Wang, R. and Zitikis, R. (2021). An axiomatic foundation for the expected shortfall.  \emph{Management Science}, \textbf{67}(3), 1413--1429.
	
	


    \bibitem[\protect\citeauthoryear{Armenti et~al.}{Armenti et~al.}{2018}]{armenti2018multivariate}Armenti, Y., Crépey, S., Drapeau, S., and Papapantoleon, A. (2018). Multivariate shortfall risk allocation and systemic risk. \emph{SIAM Journal on Financial Mathematics}, \textbf{9}(1), 90--126.

   \bibitem[{\protect\citeauthoryear{R\"{u}schendorf}{2006}}]{ruschendorf2006law}R\"{u}schendorf, L. (2006). Law invariant convex risk measures for portfolio vectors. \emph{Statistics \& Risk Modeling}, \textbf{24}(1), 97--108.

   \bibitem[{\protect\citeauthoryear{Burgert and R\"{u}schendorf}{2006}}]{burgert2006consistent}Burgert, C. and R\"{u}schendorf, L. (2006). Consistent risk measures for portfolio vectors. \emph{Insurance: Mathematics and Economics}, \textbf{38}(2), 289--297.
   
\end{thebibliography}

\end{document}